\documentclass[graybox]{svmult}

\usepackage{mathptmx}       
\usepackage{helvet}         
\usepackage{courier}        
\usepackage{type1cm}        

\usepackage{makeidx}         
\usepackage{graphicx}        
\usepackage{multicol}        
\usepackage[bottom]{footmisc}
\usepackage{amssymb}    

\makeindex             

\newcommand{\kms}{km s$^{-1}$}

\newcommand{\cmN}{cm$^{-2}$}
\newcommand{\cmn}{cm$^{-3}$}
\newcommand{\lam}{$\lambda$}
\newcommand{\arcsec}{$^{\prime\prime}$}  
\begin{document}

\title*{Physics of the Inner Ejecta}
\titlerunning{Physics of Inner Ejecta} 
\author{Fred Hamann}
\institute{Department of Astronomy, University of Florida, 
   211 Bryant Space Science Center, Gainesville, FL 32611-2055 
   \email{hamann@astro.ufl.edu} }  
\authorrunning{Hamann} 
%
%
\maketitle

\abstract{
Eta Carinae's inner ejecta are dominated observationally 
by the bright Weigelt blobs and their famously rich spectra of 
nebular emission and absorption lines. 
They are dense ($n_e\sim 10^7$ to 
$10^8$ \cmn ), warm ($T_e\sim 6000$ to 7000 K) and slow moving 
($\sim$40 \kms ) condensations of mostly neutral (H$^0$) gas. 
Located within 1000 AU of the central star, they contain heavily 
CNO-processed material that was ejected from the star about  
a century ago. Outside the blobs, the inner ejecta include 
absorption-line clouds with similar conditions, 
plus emission-line gas that has generally lower 
densities and a wider range of speeds (reaching a few 
hundred \kms ) compared to the blobs. The blobs appear to 
contain a negligible amount of dust and 
have a nearly dust-free view of the central source, 
but our view across the inner ejecta is severely affected 
by uncertain amounts of dust having a patchy distribution 
in the foreground.  Emission lines from the 
inner ejecta are powered by photoionization and fluorescent 
processes. The variable nature of this emission, occurring 
in a 5.54 yr  ``event" cycle, requires specific changes to the 
incident flux that hold important clues to the nature of 
the central object.} 


\section{Introduction}  

The ``inner ejecta'' of $\eta$ Car reside in the bright  
core of the Homunculus Nebula, spatially unresolved in seeing-limited 
ground-based images.  Spectra show a complex amalgam 
of features, including broad emission lines from the stellar wind and 
a vast number of narrower lines from the 
ejecta \cite{gav53,ADT53,AD66,HA92,HDJ94,dam98}.  High-resolution images 
using speckle techniques \cite{WEI86,HOF88}, and later the Hubble Space 
Telescope {\it (HST)\/} \cite{WEI95, ebb94, mors98, s+04},  revealed 
several bright objects less than 0.3{\arcsec}  apart, customarily 
labeled A, B, C, D.   The first {\it HST\/} spectra showed that A, the 
brightest object, is the central star, while the others -- the 
``Weigelt knots'' or "Weigelt blobs" -- are slow-moving nebular ejecta 
that produce strong narrow emission lines while also reflecting the 
star's light \cite{kd95,kd97}.   Their origin has not been explained, 
and this article is concerned mainly with their present-day nature.

The blobs appear to be located near the Homunculus' midplane, which is 
usually assumed to lie close to the star's equatorial plane and  
the orbital plane of the binary \cite{kdrh97}.  They are on the near 
side, moving away from the star at speeds of 30 to 50 km s$^{-1}$ -- 
less than a tenth as fast as the Homunculus lobes \cite{kd97, zeth99, niel07} 
Ejection dates based on proper motions have ranged from 1890 to 1940,  
well after Great Eruption in the 1840s \cite{WEI95,kd97,s+04,dor04}.\footnote{
   In principle, long-term acceleration might affect 
   this question \cite{kd97,s+04}.}   
The Weigelt Knots are most often linked with the ``Little Homunculus''  
ejected  during the second eruption in the 1890's 
(\cite{Bish03,Bish05}; see chapters by Weigelt and Kraus and by Smith 
in this volume).

At present, objects B, C and D were located 0.1{\arcsec} to 0.3{\arcsec} 
northwest of the star, corresponding to 300--1000 AU in deprojected 
distance or a light travel time of several days.\footnote{
     $D = 2300$ pc for $\eta$ Car, see 
     chapters in this volume by Humphreys and Martin 
     and by Walborn.}  
Their apparent sizes are somewhat less than 0.1{\arcsec} or $\sim 200$ AU, 
but these are just 
the brightest peaks in a complex pattern of emission and reflection (affected by 
extinction) that extends out to 0.4{\arcsec} or more from the central star. 
Little is known about the fainter associated emission/reflection regions, 
but altogether we call this ensemble of nebular material the 
``inner ejecta.''\footnote{  
    Additional fainter knots are noted in some papers.  One must 
    be wary, however, because the HST's optical point spread function 
    has a ``ring of beads'' which is not entirely removed by 
    deconvolution using the standard STScI software. }

Spectroscopic studies of the inner ejecta have pursued three 
main goals. The most basic is to estimate physical properties:   
density, temperature, ionization, kinematics, composition, 
and mass. These parameters may be  
clues to the nature and history of the central star(s). 
Another goal is 
to characterize the spectrum of the central source, 
e.g., as a binary system, by considering the nebular gas as a 
light reprocessing machine. The excitation, photoionization 
and specific emissions from the gas depend on illumination by 
the central source in 
the UV and unobservable far-UV. These are critical wavelengths 
for testing models of the central star or stars.  Finally, a third goal is 
to use the exceptionally bright and rich (and sometimes 
very unusual) line emission to study basic atomic physics 
and line formation processes.

In this review we focus mostly on spectroscopy of the brightest Weigelt 
knots B, C and D, including a new analysis of D based on {\it HST} 
Treasury spectra obtained in 2002--2003. We also briefly discuss 
two other phenomena, namely, narrow nebular absorption lines that 
appear throughout the inner ejecta \cite{stis99,gull01, gull05, gull06,
niel07} and a remarkable emission line region known as the 
``Strontium filament" \cite{gull01, zeth01sr}. 
Strictly speaking these lie outside the inner ejecta
as defined above, but they also provide insights into 
the nature of the central star and inner ejecta.  

One important uncertainty is the nature of localized dust extinction. 
We know that the dust around $\eta$ Car is patchy on small scales, 
in order to explain the blobs' high apparent 
brightness relative to the star.  Our line of sight to the latter 
has several magnitudes more extinction than B, C, and D
which are less than 0.3{\arcsec}  away \cite{kd86, HA92, kd95, HGN06} -- 
or at least this was true a few years ago \cite{MDK06}. Moreover,  
the visual-wavelength Weigelt blobs appear almost inversely correlated 
with the spatial distribution of mid-IR (warm) emission by dust.  
These factors lead to a fundamental ambiguity about whether the 
observed blobs are distinct gas condensations or simply minima in the 
intervening dust. Perhaps they are a combination of both.

Another issue to keep in mind is the overall transience of 
the inner ejecta. The Weigelt blobs were ejected from 
the star less than $\sim \, 120$ years 
ago. If the blobs are not confined by surrounding pressures, 
they should expand and dissipate in roughly a sound-crossing 
time, on the order of 75 years.  Their spatial and spectral  
appearance has changed in recent years, see chapter by 
Humphreys and Martin in this volume.  
High-ionization emission lines such as  He~I, which  
are now trademarks of the knot spectra, did not appear 
until the  1940's \cite{RMH08,fea01}.  Continuing changes 
will occur as the material expands and moves farther from the star.  
Meanwhile there are cyclical changes with a 5.54-year 
period, usually attributed to the binarity of the central object 
(\S2 below). Any discussion of the inner ejecta must be framed 
with reference to the epoch of the observations.


\section{Spectroscopic Overview of the Weigelt Blobs}  

The Weigelt blobs produce H~I and He~I recombination lines and more than 
2000 other identified emission lines spanning a range of ionizations 
from Ca$^+$ and Ti$^+$ up to S$^{+2}$, Ar$^{+2}$, and Ne$^{+2}$. 
Most of the UV, visual, and near-IR features belong to singly-ionized 
iron group species, notably Fe~II and [Fe~II]. Extensive line lists are 
available \cite{viot89, HA92, HDJ94, dam98, wall01, zeth01cr, 
vern02}.  A particularly interesting aspect is the variety of strong 
fluorescent lines, whose upper energy states are vastly overpopulated by 
photoexcitation because of accidental wavelength coincidences with the 
H~I Lyman series or other strong lines.  Fluorescent  Fe~II 
$\lambda\lambda$2507,2509 are the strongest emission features in 
near-UV spectra;  their enhancements compared to other Fe~II lines 
are larger in $\eta$ Car than in any other known object 
\cite{kd95,viot89,joha99,joha04}.                                      
Altogether, the varieties of lines and excitation processes provide 
a broad array of diagnostics with which to study both the blobs and the central 
object.

An obstacle to these studies has been the seeing-limited angular resolution 
of ground-based spectroscopy,  typically $\sim$ 1{\arcsec}.  To some extent 
the line profiles distinguish between blobs and the stellar wind;  velocity 
dispersions are 40--70 km s$^{-1}$ vs.\ several hundred km s$^{-1}$ 
respectively.   However, the rich  
narrow-line spectrum has complex blends that resemble broad features,  
{\it HST\/} spectra of the inner ejecta show differences 
at sub-arcsec scales, and reflection  and projection effects occur 
at all scales.  Moreover, some high ionization forbidden lines originate 
in high velocity gas, distinct from the blobs and not directly part of the 
stellar wind (\S4.1).   Spatial resolution better than 0.2{\arcsec} is 
therefore essential for detailed studies.

The spectra of the stellar wind and the blobs vary with a 5.54 yr period  
\cite{zane84, whit94,dam96,dam08b, dam08a, MK04, meh10a}. This cycle is 
punctuated by ``events'' defined by the disappearance of high-ionization 
emission from the blobs and inner ejecta, notably [Ne~III], [Fe~III], 
[Ar~III] and He~I.  These features vanish on time scales of 
1--6 weeks and then recover more slowly afterward.  
Meanwhile other phenomena occur, including an abrupt drop in the 
2--10 keV X-ray emission from the colliding winds 
\cite{ICD99, CIS01,corc05, henl08}.\footnote{
     See chapter by Corcoran and Ishibashi in this volume.}
A major goal of $\eta$ Car studies since 1997 has been to understand 
the spectroscopic events, which must be  
linked to some basic aspect of the central source.  The disappearance 
of high-ionization [Ne~III], [Fe~III], [Ar~III] and He~I lines 
is almost certainly caused by an abrupt drop in the far-UV flux 
incident on the blobs and inner ejecta as originally proposed 
by \cite{zane84} (see \S3 and \S5 below).  What physical effects 
in the central object can change its spectral energy output?

The regularity of the event cycle is usually interpreted as evidence 
that the central object is a 5.54 yr binary \cite{dam96, dam08a,dam00}),  
but no specific model has emerged that explains the full range 
of phenomena (see \cite{meh11,RMH08,dam08a} for recent discussions). 
The proposed companion star is less luminous but hotter than the very 
massive primary \cite{meh10a}.  It contributes most of the helium-ionizing 
far UV flux and thus controls the high-ionization emission lines.  The 
orbit is highly eccentric, so close interactions occur only for a brief 
time near periastron.  Between spectroscopic events, the stars are widely 
separated and both contribute to the ionization and excitation of the 
inner ejecta. During an event, near periastron, the hot companion 
plunges deep inside the dense primary wind, so its contribution  
to the emergent far-UV emission is briefly obscured.\footnote{    
     (Editors' comment:)  Strictly speaking, the disappearance  
     of far-UV near periastron may be caused by mass accretion 
     onto the secondary star as proposed by Soker et al.  See 
     \cite{meh11}, the chapter by Davidson in this volume, and 
     references therein.}  
The inner ejecta then become less ionized because they receive light 
from only the cooler primary, whose wind structure and spectral energy output 
might also be altered by the binary encounter \cite{kd99,s+03lat,kd05,dam08a}. 
A model of this type can account for the X-ray variations 
\cite{ICD99,pc02,corc05}.
But there are many uncertainties, including the nature of the hot 
companion star, the orbit parameters, and properties of both winds.  
Any model of the binary system must explain the spectroscopic 
properties of the inner ejecta and, specifically, the changes that occur 
throughout the 5.54 yr cycle.  See comments and references in the 
chapter by Davidson in this volume.  

                                                                 
\section{Blob D and the 2003.5 Spectroscopic Event}  

The spectroscopic event that occurred in mid-2003 was studied 
at wavelengths ranging from radio through X-rays.\footnote{
    See other chapters in this volume as well as \cite{kd05,heii06,dam08a} 
    and many refs.\ therein.}   
Here we summarize key results for the inner ejecta, especially the 
almost-resolved ($\sim$ 0.1{\arcsec}) spectra of blob D that were obtained 
as part of the {\it HST\/}  Treasury program on $\eta$ Car.\footnote{    
    http://etacar.umn.edu/. } 
This publicly available dataset provides the most complete and reliable 
existing information of this type. 
The 2003.5 HST observations covered roughly 1630 \AA\ to 10100 \AA\ at 
resolution $\sim$ 40 \kms and $\sim$ 0.1{\arcsec};  no better data on 
an event are expected in the forseeable future.\footnote{   
     The only instrument with adequate spatial resolution, HST/STIS, 
     was inoperative during the subsequent event in 2009.  Moreover, 
     the star's rising brightness progressively makes the Weigelt 
     blobs harder to observe \cite{MDK06}.  Thus it is very conceivable 
     that no one will ever obtain new ``event'' spectra of 
     these objects as good as the 2003.5 STIS data.}

Figures 1--8 illustrate spectral properties of blob D measured on six 
occasions across the 2003.5 event. The line with highest ionization, 
[Ne~III] $\lambda$3868,  showed the most rapid decline and complete 
disappearance. Figure 1 shows it along with the more complex 
Balmer H8.  The narrow in situ H~I emission disappeared 
while the reflected broad stellar wind component weakened in 
emission but strengthened in its blueshifted P Cyg absorption (see also 
\cite{kd05,HGN06}).  Narrow lines with slightly lesser ionization, e.g., 
[Ar~III] $\lambda$7135 and [S~III] $\lambda$9532, disappeared about  
2--3 weeks later than [Ne~III] (see Figure 2 in \cite{hama05}).    At 
lower ionizations, features like [N~II] $\lambda$5755 
lagged even farther, cf.\ Figures 1 and 2.  At $t = 2003.38$ the event 
was significantly underway in [Ne~III] but not in [N~II], and at 
2003.48 the [Ne~III] line had disappeared while [N~II] emission was 
still present. 

\begin{figure}[ht]   
\center
\includegraphics[scale=0.6]{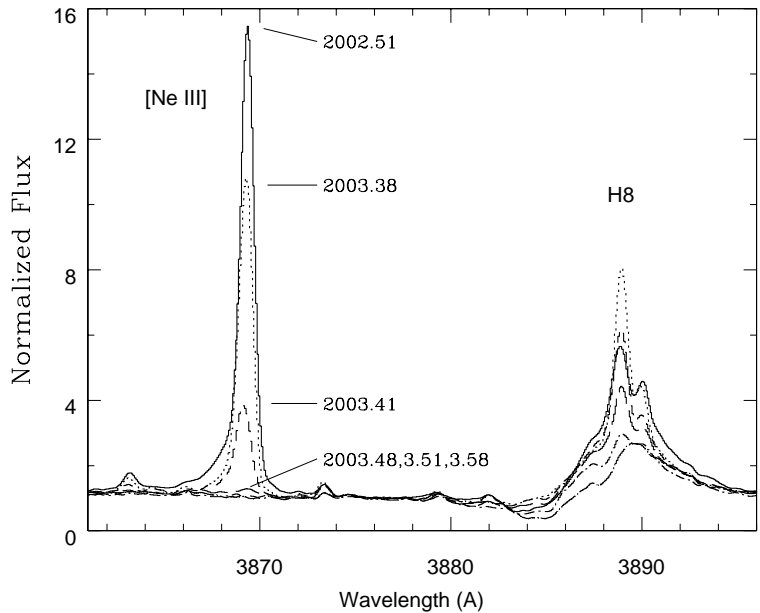}
\caption{[Ne~III] $\lambda$3868 and H8 emission lines in
 blob D on the dates 2002.51 (solid line), 2003.38 (dotted),
 2003.41 (short dash), 2003.48 (long dash), 2003.51 (short
 dash-dot), 2003.58 (long dash-dot). The [Ne~III] peak heights are
 marked on the different dates. The weak emission bump at 3863.2 \AA\ 
 is Si~II 3863.69 \AA\.  The broad P Cygni profile in H8 is reflected 
 starlight. From \cite{hama05}.}
\end{figure}

\begin{figure}[ht]   
\center
\includegraphics[scale=0.62]{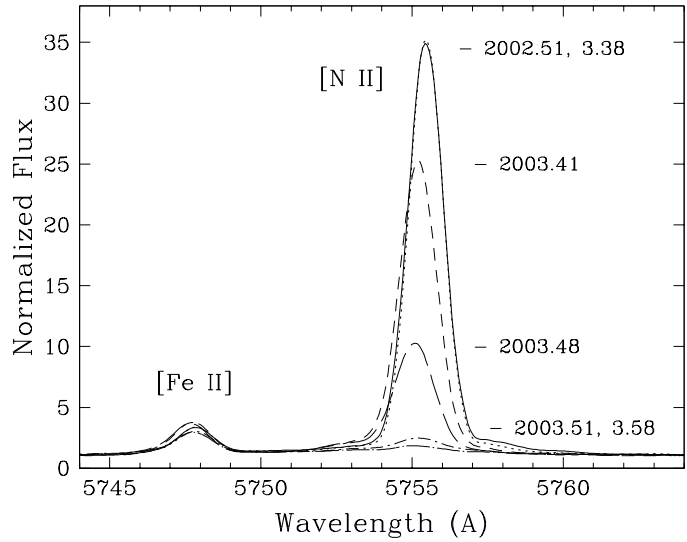}
\caption{[N~II] $\lambda$5755 and [Fe~II] $\lambda$5747 measured 
 in blob D during the 2003.5 spectroscopic event. The labels 
 and line styles for the different dates match Figure 1. 
 }
\end{figure}

Continuing toward lower ionization, the behavior of Fe~II, [Fe~II],
and [Ni~II] ranged from modest weakening to modest {\it strengthening\/} 
during the event. The lowest ionization lines measured in blob D, 
[Ca~II], Ti II and V II, all became stronger (see also \cite{dam98}). 
Figure 3 shows that [Ca~II] $\lambda$7291 and $\lambda$7323 lines   
nearly tripled in strength, while [Ni~II] $\lambda$7303 mildly 
strengthened and He~I $\lambda$7281 disappeared.  Some Ti~II 
lines (not shown) approximately doubled in strength.   

\begin{figure}[ht]    
\center
\includegraphics[scale=0.6]{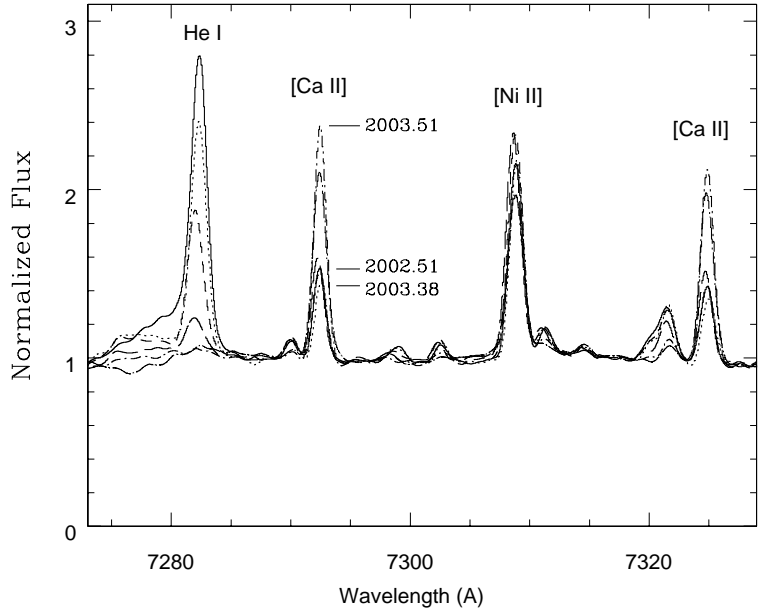}
\caption{[Ca~II] $\lambda$7291,7323, [Ni~II] $\lambda$7307, and
 He~I $\lambda$7281 in blob D during the 2003.5 event. 
 The [Ca~II] lines tripled in strength, the [Ni~II] 
 strengthened slightly, and 
 the narrow He~I line became $>$30 times weaker during the 2003.5 
 event. Labels and line styles match Figure 1. From \cite{hama05}}
\end{figure}

These phenomena were correlated with the ionization energy needed to   
create each ion.   Ne$^{+2}$, Ar$^+$, S$^{+2}$ and N$^+$ require 
41.0, 27.6, 23.3 and 14.5 eV, respectively,  
while Fe$^+$ and Ni$^+$ require 7.9 and 7.6 eV,  and Ti$^+$ and 
Ca$^+$ need just 6.8 and 6.1 eV.  The narrow in situ (not reflected) 
emission lines of H~I and He~I are consistent with this pattern if   
they form, as expected, by recombination in regions of H$^+$ (13.6 eV) 
and He$^+$ (24.6 eV).  He~I lines behaved approximately like 
[Ar~III] and [S~III], while the H~I lines declined like [N~II] 
(see also \cite{hama05}). 

The only obvious exceptions to this simple ionization trend were  
fluorescent features excited by H~I Lyman lines.  Figures 4 and 5 show 
the dramatic weakening of fluorescent O~I $\lambda$8446 pumped by 
Ly$\beta$, and Fe~II lines excited by Ly$\alpha$.   Their behavior 
differed from the other Fe~II and [Fe~II] lines and seems to contradict 
the ionization trend.  However, their emission depends not only on the 
amounts of Fe$^+$ and O$^0$,  but also on H$^+$ which produces Ly$\alpha$ 
and Ly$\beta$ by recombination (see \S4.6).  In fact, the fluorescent lines  
behaved much like the narrow Balmer and Paschen lines in blob D, 
consistent with the ionization trend.  This can be seen in 
Figure 4, which shows the disappearance of narrow H~I Pa15 along 
with the fluorescent Fe~II and O~I features. All these lines 
disappeared when the gas became mostly H$^0$.     

\begin{figure}[ht]    
\center
\includegraphics[scale=0.6]{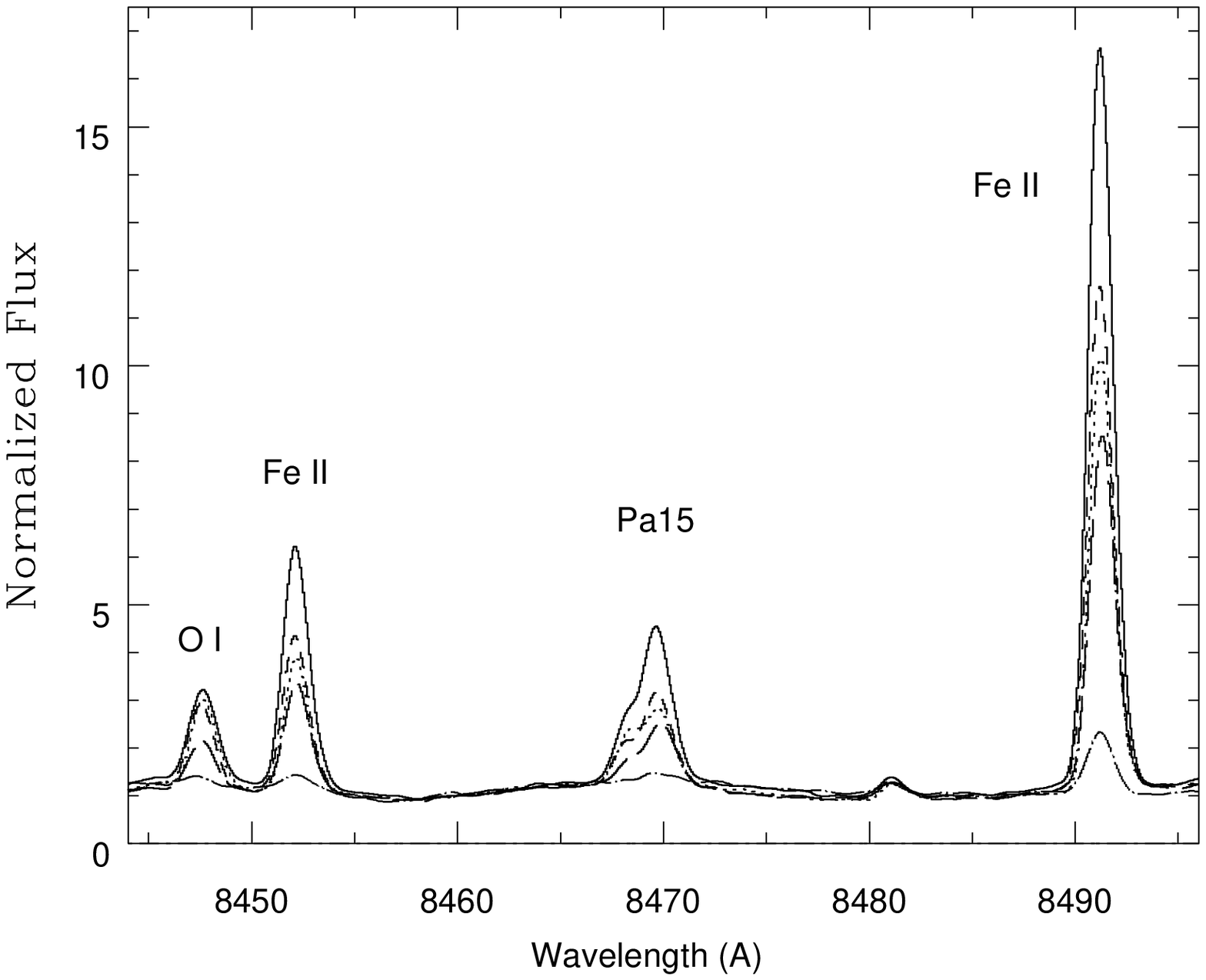}
\caption{H~I Pa15 and the fluorescent lines O~I $\lambda$8446 
 pumped by Ly$\beta$ and Fe~II $\lambda$8451 and 
 $\lambda$8490 pumped by Ly$\alpha$ in blob D. The different line 
 styles represent different observation dates as in Figure 1. 
 The fluorescent lines decreased much more dramatically during 
 the event than other collisionally-excited lines of Fe~II or [Fe~II].
 From \cite{hama05}.}
\end{figure}

\begin{figure}[ht]    
\center
\includegraphics[scale=0.62]{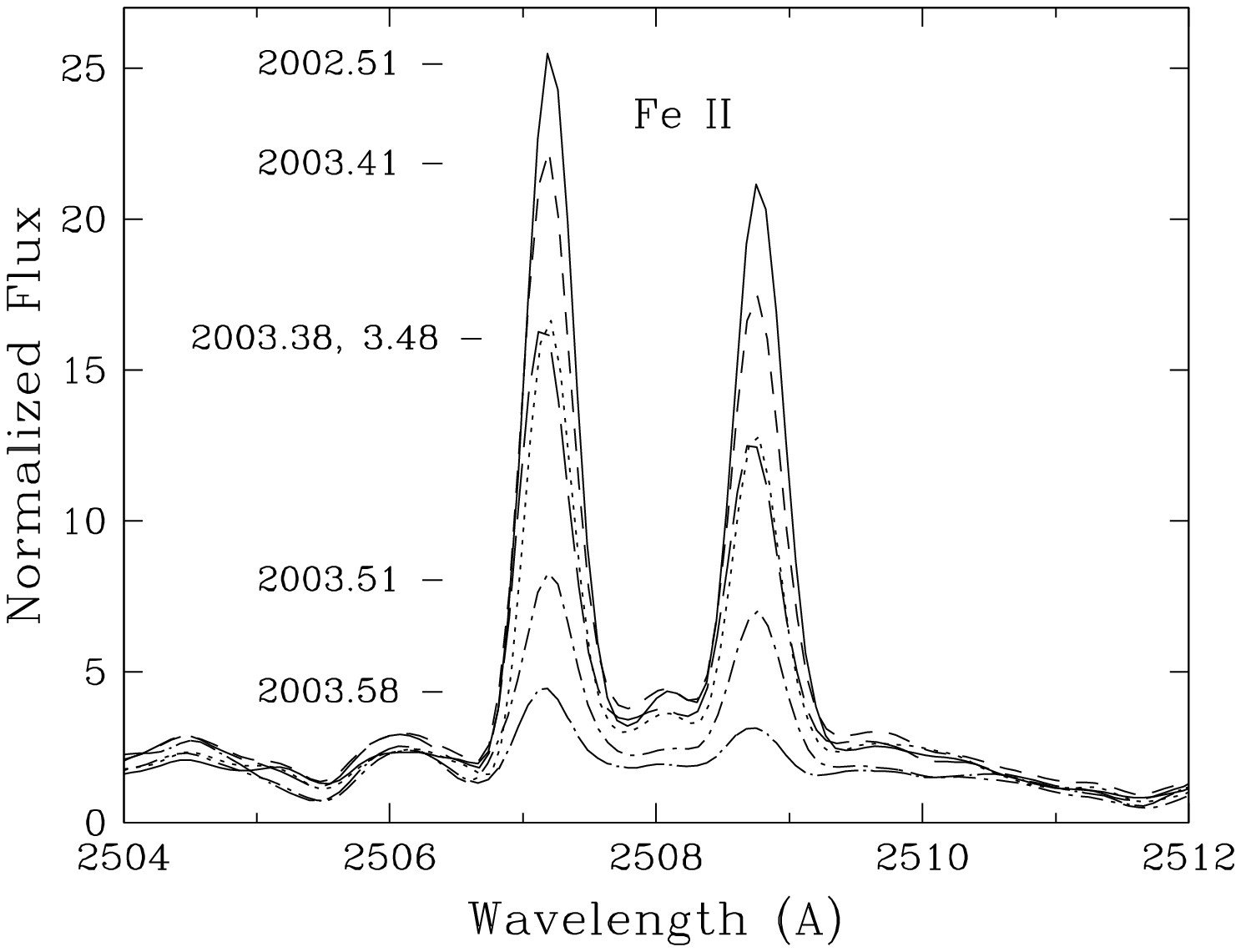}
\caption{The fluorescent lines Fe~II $\lambda$2508 and 
$\lambda$2509 pumped by Ly$\alpha$ in blob D. The line labels 
 and styles match Figure 1.}
\end{figure}

Figure 6 summarizes the ionization trend for narrow lines in blob D 
during the 2003.5 event.    Damineli et al.\ \cite{dam08a,dam08b} 
produced similar plots based on ground-based observations 
of a spatially unresolved mixture of regions, but with 
better temporal sampling and a longer temporal baseline.  They show that 
the high-ionization lines disappeared abruptly, in just 5 to 10 days.  
The highest ionization line, [Ne~III] $\lambda$3868, was extinguished 
first, followed by He~I $\lambda$6678 about 5 days later, then 
[S~III] $\lambda$6312 6.5 days later, and finally [N~II] $\lambda$5755 
8.5 days after that. The highest ionization lines were also the last to 
recover after the event, roughly in reverse order of their disappearance;  
the recovery times were more gradual (months) than the disappearance 
times (days to weeks).

The shortest disappearance times might be limited by recombination 
rates, which depend on gas density  (see \S4.2). But this is unlikely 
to dominate the other temporal behaviors,  since the recovery times 
(when overall ionization is increasing) are much 
longer.   Therefore, the emission line changes probably trace the 
central source's spectral changes as viewed by the inner ejecta. 
In particular, the far-UV flux that regulates [Ne~III] emission 
must have been extinguished faster, earlier and then recovered  more slowly, 
than the lower energy spectrum that controls the lower ions. The incident 
photon energy distribution must have progressively ``softened'' until 
mid-event when all of the far-UV was gone. Then the spectrum hardened 
again as the far-UV recovered over a period of months 
(see also \cite{dam08b,meh10a,meh11} and \S5 below).  

\begin{figure}[ht]    
\center
\includegraphics[scale=0.6]{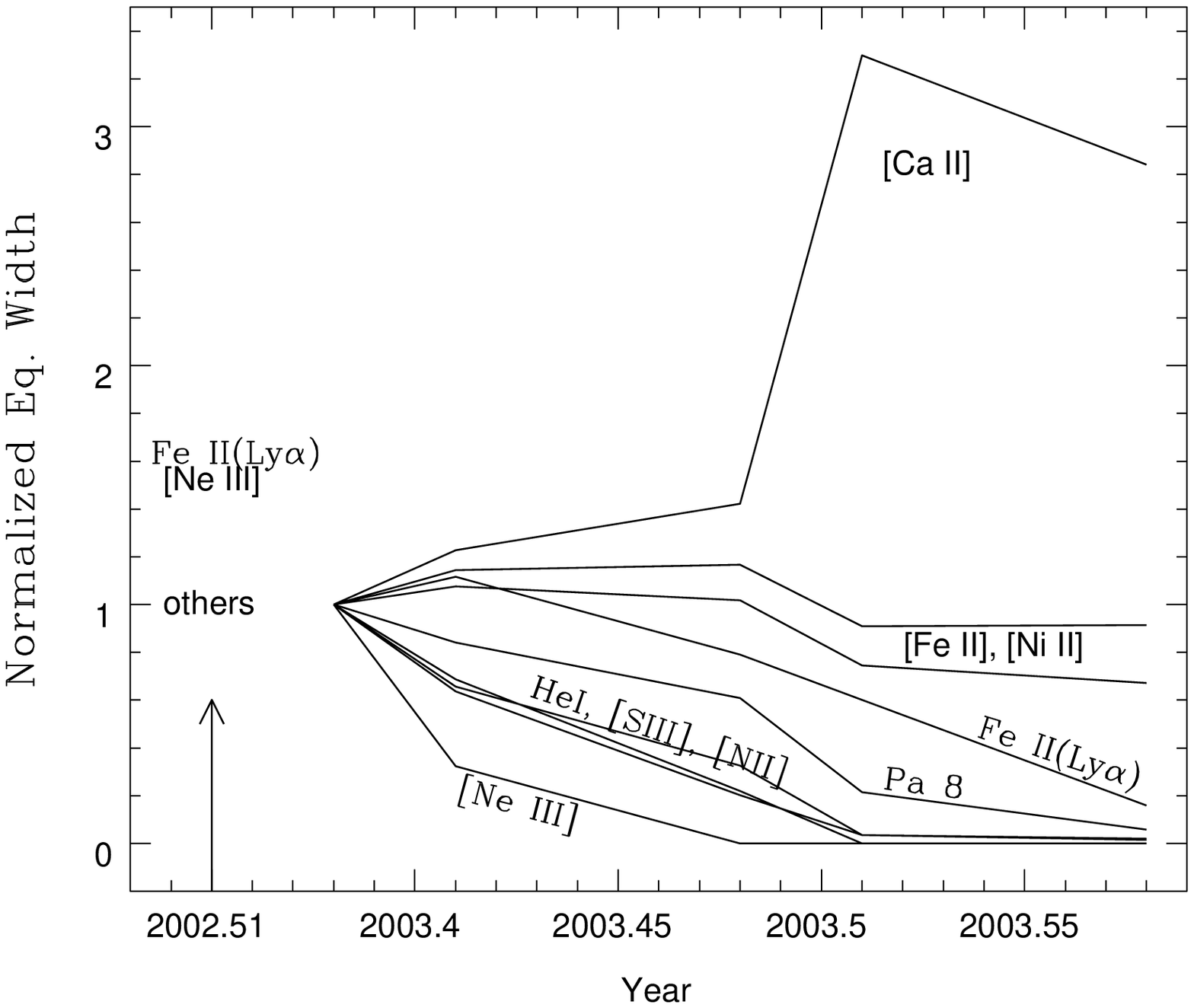}
\caption{Measured equivalent widths (\AA ) of representative 
 lines in blob D, normalized to their value in 2003.38. 
 Fe~II(Ly$\alpha$) represents an average of several 
 fluorescent Fe~II lines pumped 
 by Ly$\alpha$. The time scale is distorted for the 2002.51 data, 
 Vertical positions of labels for 2002.51 indicate 
 equivalent widths relative to 2003.38, where ``others'' 
 refers to all lines plotted here except [Ne~III] and 
 Fe~II(Ly$\alpha$). Higher ionization lines weakened sooner and 
 more completely during the event, while the lowest ionization 
 lines of [Ca~II] became stronger. From \cite{hama05}.}
\end{figure}

Another spectral change tied to the ionization was the weakening of 
narrow Balmer {\it absorption\/} during the 2003.5 event. 
Unlike the stellar wind's P Cyg absorption, these lines are formed 
in regions far outside the wind and even outside the Weigelt 
blobs (\S4.6 below).   They require a large column density of 
dense, partially ionized gas with a significant population of H$^0$ 
in the $n=2$ level.  Figure 7 shows the disappearance of H$\gamma$ 
and H$\delta$ absorption  in blob D during the 2003.5 event. The 
corresponding H$\beta$ feature (not shown) weakened by a factor of 
$\sim$ 2 but did not disappear. This narrow absorption line behavior 
did not exactly follow the weakening of the narrow H~I emission lines 
or the Fe~II emission lines pumped by Ly$\alpha$, but a decline 
during the 2003.5 event surely did occur in blob D. 
Weaker Balmer absorption strength indicates a drop in the H$^0$ 
$n=2$ population, related to a lower degree of 
ionization.\footnote{
    Johansson et al. \cite{joha05} report that corresponding narrow 
    absorption in H$\alpha$ {\it strengthened} during the 2003.5 event, 
    and the same is true of ground-based spectra of the star plus 
    ejecta in the 2009.0 event \cite{rich2010}. We cannot make 
    direct comparisons to these results, because the spatial 
    coverage was different and instrumental saturation 
    in the {\it HST} data may have degraded spectral extractions 
    near the peak of H$\alpha$.   We can only speculate that 
    the ground-based H$\alpha$ absorption results were 
    affected by blending with the narrow emission from the blobs, which 
    is time variable and stronger in H$\alpha$ than in the other  
    Balmer lines.  H$\alpha$ might also be less sensitive to changes in 
    the $n=2$ population if its larger oscillator strength leads to greater 
    line saturation compared to H$\gamma$ and H$\delta$.  } 
We conclude that the absorbing gas participated in a spectroscopic/ionization 
event similar to the emission line regions in the inner ejecta.

\begin{figure}[ht]    
\center
\includegraphics[scale=0.62]{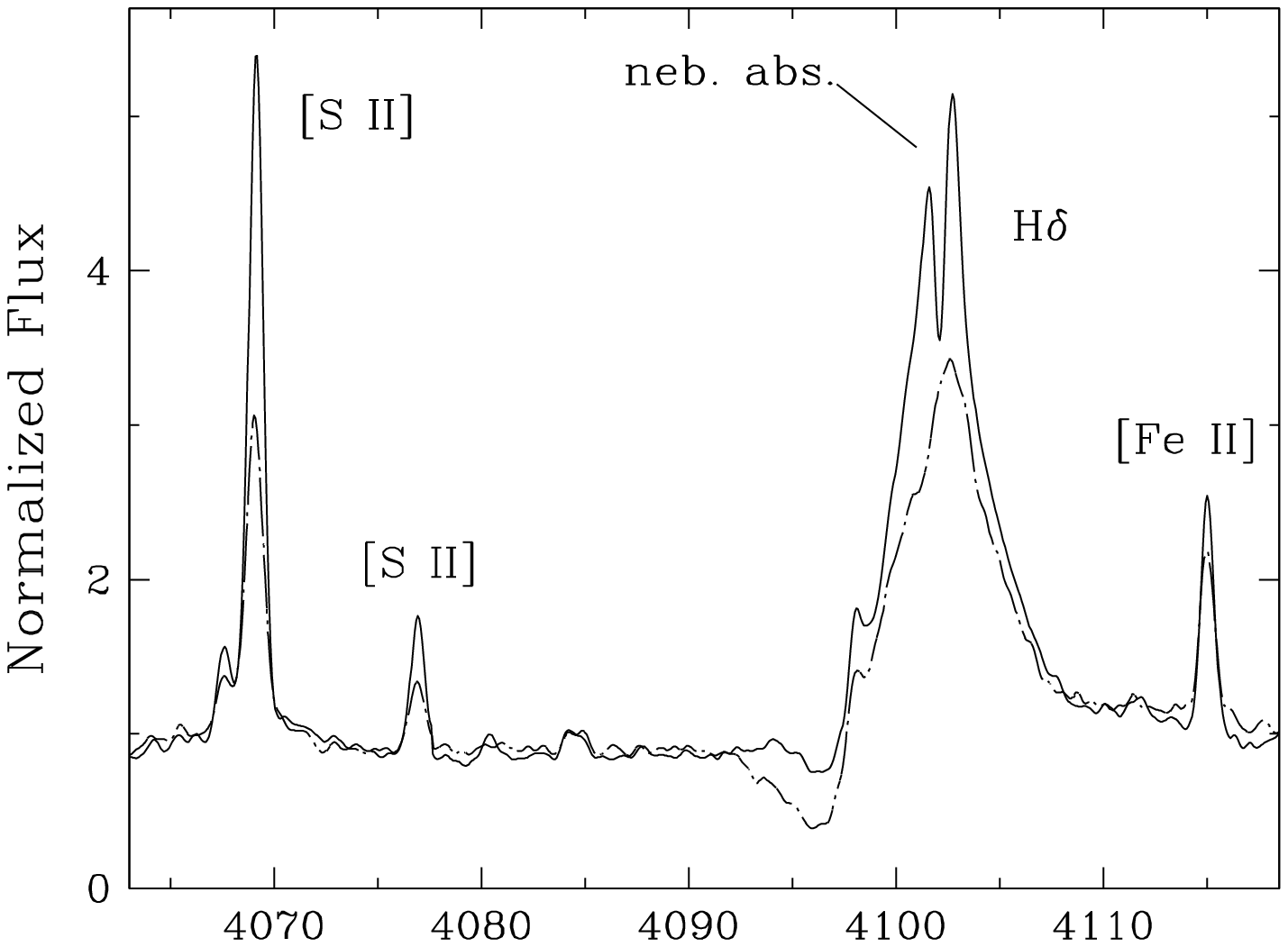}
\includegraphics[scale=0.62]{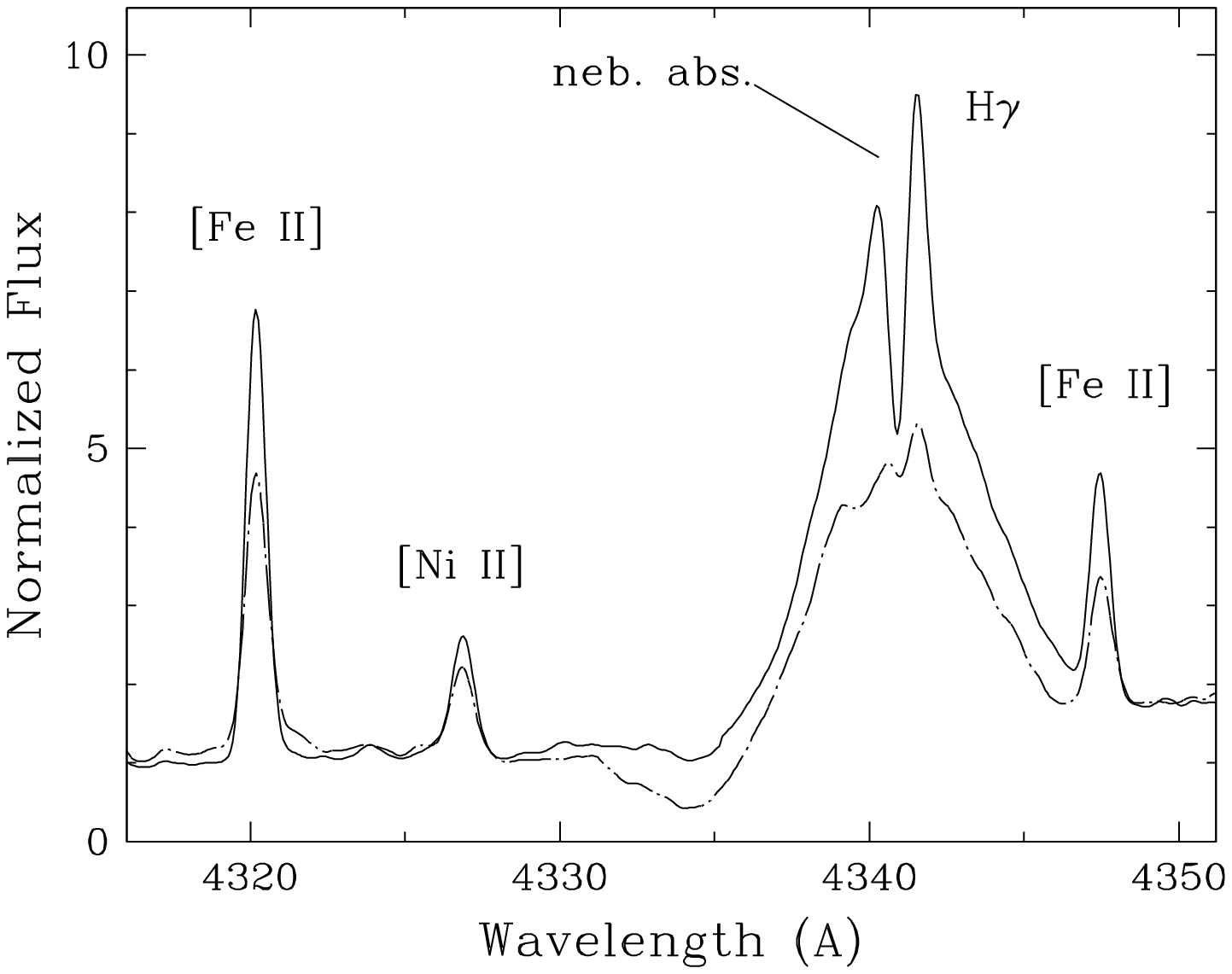}
\caption{The H$\delta$ (upper panel) and H$\gamma$ (lower panel) 
lines are plotted along with various lines of [Fe~II], [S~II] 
 and [Ni~II] measured in blob D on the dates 2002.51 (solid line) 
 and 2003.58 (long dash-dot). 
 The P Cygni shaped Balmer lines are seen in reflected light 
 from the star. Their broad emission weakened and broad absorption 
 strengthened during the event (2003.58). Narrow nebular 
 absorption (neb. abs.) in both H$\delta$ and H$\gamma$, at roughly 
 $-50$ \kms\ heliocentric, disappeared during the event.   
 }                  
\end{figure}

Closer inspection of these data 
suggests that the temperature in blob D also fell by 
a moderate amount during the 2003.5 event.   For each ion 
species, emission lines arising from higher energy states generally 
faded more dramatically.   Figures 7 and 8 show, for example, that 
[S~II] $\lambda$4069 and $\lambda$4076 declined by a factor of two 
during the event while [S~II] $\lambda$6716 and $\lambda$6731 
decreased by only $\sim$10\%.  These lines are collisionally excited, 
and the upper states of $\lambda\lambda$4069,4076 and 
$\lambda\lambda$6716,6731 have energies of 3.0 and 1.8 eV 
repectively.  There is also a density dependence \cite{oste06,hama94}, 
but if we assume that the density did not change much during the  
event, then the observed change in the line ratio indicates 
a drop in temperature. We cannot derive a specific temperature without 
knowledge of the density, but if the temperature in the S$^+$ gas was 
$\sim$7000 K before the event (2002.51), then during the event (2003.58) 
it declined by roughly 900 K (or $\sim$700 K if the initial 
temperature was 6000 K).  
Similar effects can be seen in [Fe~II] and [Ni~II], see \S4.2 below.

\begin{figure}[ht]   
\center
\includegraphics[scale=0.62]{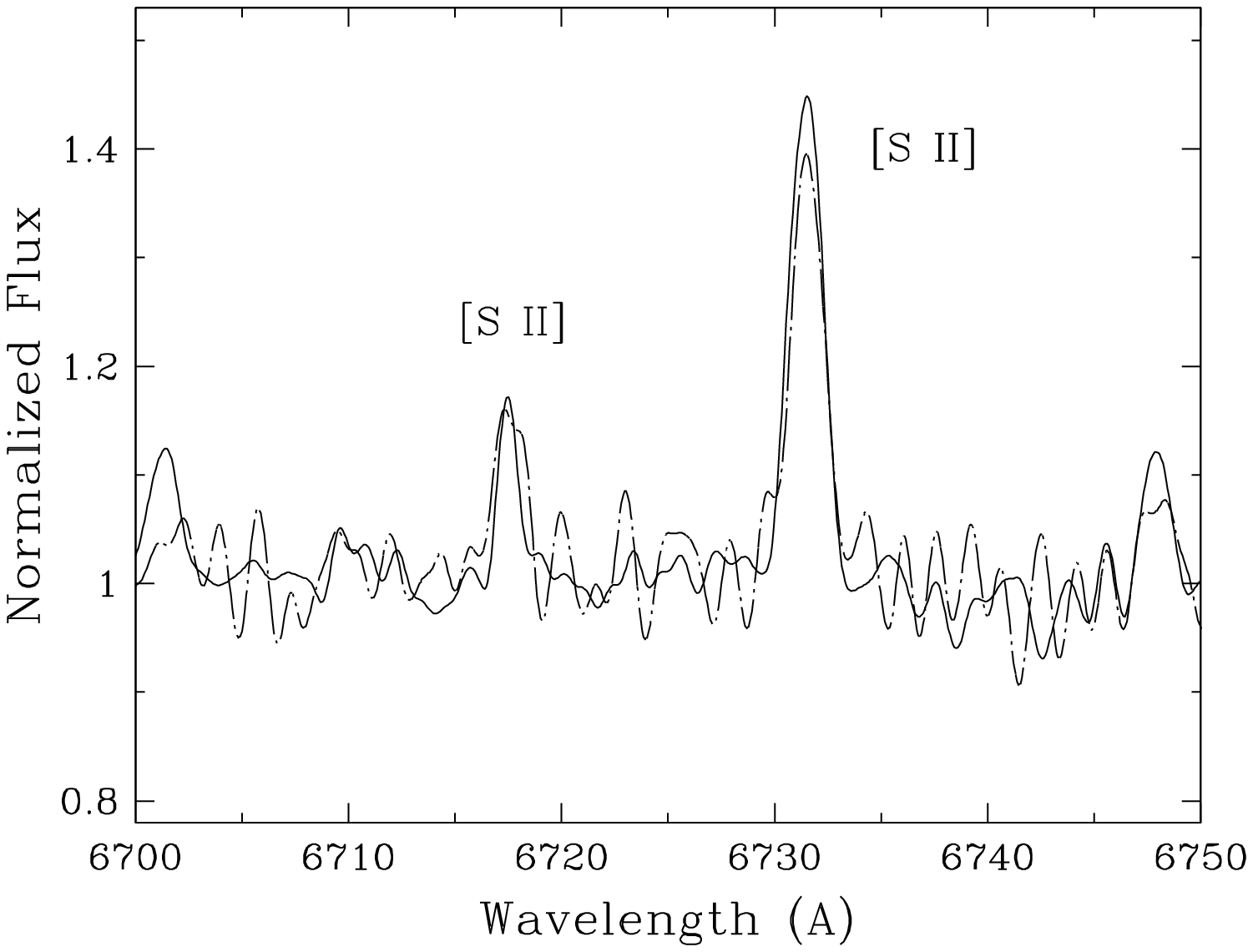}
 \caption{[S~II] lines measured in blob D in 2002.51 
 (solid line) and 2003.58 (log dash-dot) are plotted 
 for comparison to the higher energy [S~II] doublet shown in 
 Figure 7. The different flux changes between these lines suggest 
 that the temperature dropped by roughly 15\% during the event.
 }
\end{figure}

Finally, we note that the Balmer P Cygni absorption lines, 
formed in the stellar wind and reflected by dust in blob D, 
varied contemporaneously with the narrow in situ 
emission lines discussed above (Figures 1 and 7). Detailed comparisons 
have shown that these changes tracked each other 
to within a month \cite{stah05,HGN06, kd05, dam08b}.  This 
relationship indicates again that the blobs were responding 
to changes in the radiative output from the central source, 
presumably the opaque primary stellar wind plus the hot secondary 
star.  (Variations in the kinetic energy of the primary wind 
would affect the blobs only after a travel-time delay of a year 
or more.)  This result implies that ionization and excitation in 
the blobs (and probably all of the inner ejecta) are dominated 
by the radiative flux from the central star, rather than shocks 
or other mechanical processes (see \S5).


\section{Analysis: Physical Properties and Peculiarities}  

Here we review basic physical properties that can be derived from the 
spectral lines in the inner ejecta.\footnote{ 
   Some of the text in \S4.1 was contributed by 
   K.\ Davidson and A.\ Mehner. }      
We discuss blobs B, C and D interchangeably because their spectra are 
broadly alike.

\subsection{Kinematics and Location of the Highly Ionized Gas}

Kinematic data are essential for defining locations and origins 
of various types of regions in the inner ejecta.  We mentioned some 
results in \S1 and \S2, and more information 
can be found elsewhere in this volume, in chapters by 
Weigelt and Kraus and by Smith.  The Weigelt knots have Doppler 
velocities around $-40$ km s$^{-1}$ in both low and high 
excitation emission lines.   High ionization features 
such as [Ne~III] have Doppler widths of 65--70 \kms\ (FWHM), 
while [Fe~II] and non-fluorescent 
Fe~II have FWHM $\sim$ 55 \kms\  (\S3, and \cite{s+04}).  
We  expect features with disparate ionization energies to form in 
different locations (\S5, \cite{vern05,meh11}).  In one 
plausible but unproven geometry, the blobs are mostly neutral (H$^0$) gas 
with ionized layers facing the central binary system. Since the hot 
secondary star is the chief source of relevant ionizing photons,  
the ionized zone varies during the 5.54 yr orbit, leading to alternate 
appearance and disappearance of high-ionization lines.  At spatial 
resolution $\sim$ 0.05{\arcsec}, {\it HST\/} spectroscopy shows that 
locations of maximum [Ne~III] and [Fe~III] brightness approximately 
match those of the low-excitation lines \cite{meh10a}.\footnote{ 
   Some earlier authors assumed that high-ionization 
   lines originate diffusely between the star and the 
   Weigelt knots \cite{vern05}.  To some extent this may 
   be true, but the brightness peaks are located as stated 
   above.  {\it HST\/} had no imaging filters suitable for 
   isolating the pertinent spectral lines. }
Higher spatial resolution will be needed to show ionization 
stratification.  The differing line widths suggest that 
high-ionization zones are only loosely related to the low-ionization 
material, and possibly ablating from the blobs.

Ground-based spectra show blue wings of [Ne~III], [Fe~III], [Ar~III], 
and [S~III], extending to local peaks near $-380$ \kms\ 
\cite{ADT53,AD66,dam98,hama94,HA92}.  These are unrelated to  
the Weigelt knots;  {\it HST\/} spectroscopy shows that the 
blue-shifted [Fe~III] originates in a slightly elongated region 
with radius $\sim$ 0.1{\arcsec}, centered near the star 
(Figs.\ 8 and 9 in \cite{meh10a}).   The simplest guess is 
that these features come from our side of a mildly oblate region 
in the outer wind, with densities low enough to emit forbidden 
lines \cite{meh10a}.  Larger-scale locations in the  Homunculus are 
also possible, however.  Helium emission is not useful in this 
regard, because the complex He~I line profiles combine several 
distinct regions, including absorption in the wind.

Apart from the Weigelt knots, narrow high-excitation forbidden lines 
appear in {\it HST\/} spectra of the star itself \cite{meh10a}.   
Based on their small widths, low velocities, and de-excitation densities, 
they represent line-of-sight gas comparable to the Weigelt knots, not 
the stellar wind.  Given {\it HST's\/} high spatial resolution, these 
features' response to the central UV output must be closely correlated 
with our direct spectroscopic view of the star itself.\footnote  
    {Strictly speaking, our view of the dense primary wind plus 
    the hot companion star.  The Weigelt knots may differ because 
    they ``see'' the star from other directions. The primary wind 
    is not spherically symmetric, and local circumstellar extinction 
    may be both patchy and variable. }    
In fact the line-of-sight [Ne~III] and [Fe~III] intensities do vary 
systematically and non-trivially through the 5.54 yr cycle 
\cite{meh10a,dam08a}.  Their growth, broad mid-cycle maximum, and 
gradual decline seem reasonable in terms of photoionization by the 
secondary star as it moves along its orbit, but no quantitative model 
has been developed. The line-of-sight data \cite{meh10a} showed a 
conspicuous brief secondary maximum in [Ne~III] and [Fe~III] several 
months before the 2003.5 event (Figure 10).  This may have been the 
time when the orbiting secondary star was optimally located for 
photoionizing our line of sight \cite{meh11}.

Narrow Balmer absorption lines also trace the ionized nebular gas (\S4.6). 
These features appear in spectra across the central 1$^{\prime\prime}$ 
to 2$^{\prime\prime}$, encompassing the star and inner ejecta 
\cite{kd01a}.  In {\it HST} spectra of the Weigelt blobs, they have 
heliocentric velocities of roughly $-$46 to $-$50 \kms\ (Figure 7). 
These values are similar to those for the narrow high-ionization emission 
lines \cite{s+04}.  During the 2003.5 spectroscopic event, narrow 
H$\beta$ absorption weakened while corresponding H$\gamma$ and H$\delta$ 
absorption disappeared, in a manner similar to the narrow H~I emission 
lines. It therefore seems likely that these absorption lines form in 
an ionized layer that is loosely related to the blobs. 
In spectra taken along our direct line of sight to the star, there 
is additional narrow Balmer absorption at $-$146 \kms\ 
\cite{gull05}. This component of ionized absorbing gas 
seems to have no relationship to either the stellar wind or the blobs.
It probably resides elsewhere in the inner ejecta, see \S4.6.


\subsection{Reddening, Extinction \& Temperature}  

Extinction and reddening by dust can dramatically affect the observations. 
In principle we can estimate the reddening, $E_{B-V}$, by comparing 
emission lines that share  the same upper level.  If optically thin, 
then their intrinsic flux ratios are  
  \begin{equation}
  {{F_1}\over{F_2}} \ = \ {{A_{1}\lambda_2}\over{A_{2}\lambda_1}}
  \end{equation}
where $F_1$, $\lambda_1$, and $A_{1}$ are the flux, wavelength and 
decay rate for line 1, etc.  Ideally we would measure enough lines 
at different wavelengths to characterize the reddening curve.   
In practice, however, there are not enough 
well-measured lines of this type.  One could simply adopt a standard 
reddening curve for the interstellar medium (e.g.\ \cite{card89}), 
but both the cirumstellar and interstellar extinction for 
$\eta$ Car are known to be anomalously gray with $A_V/E_{B-V} > 4$ 
rather than a normal value around 3.1 \cite{kdrh97}.  Thus $A_V$ 
was probably about 7 magnitudes for the central star in 1998, but 
$E_{B-V} \sim 1$ \cite{hill01}.  Both these values appear to have 
declined since that time \cite{MDK06,meh10b}.  The blobs are thought to 
have much less extinction, in order to explain their surprisingly 
large brightnesses compared to the central object \cite{kd95}.

Hamann et al. \cite{hama99} measured suitable line pairs in HST/STIS  
spectra of blobs B+D.   They found $E_{B-V} \sim $ 0.6, 0.7, and 0.8 mag, 
respectively, based on [Fe~II] \lam 3175/\lam 5551, [Fe~II] 
\lam 3533/\lam 6355, and [Ni~II] \lam 4326/\lam 7256.   Investigations 
with non-[Fe~II] lines, however, have given $E_{B-V} \lesssim 0.2$ 
\cite{vern02,meh10a}.  It is difficult to reconcile all these results,
except to note that they span different wavelengths, and that a STIS 
instrumental effect tends to counteract reddening for 
$\lambda \gtrsim 4500$ {\AA} \cite{meh10a}.

Another way to estimate both reddening and temperature is 
to compare the entire rich spectrum of [Fe~II] lines to theoretical 
predictions.  The observed [Fe~II] lines arise from metastable states 
at energies $\lesssim$ 4 eV, and densities in the Weigelt blobs 
(\S4.3 below) exceed the collisional de-excitation values for most of them.    
Thus we can reasonably assume that the level populations are close to 
local thermodynamic equilibrium.  In that case the relative [Fe~II] 
line strengths are given by Equation 1 multiplied by a Boltzmann factor 
for the upper states' populations. With enough lines spanning a range 
of wavelengths and excitation energies, we can solve for both reddening 
and temperature.  Resulting temperatures are useful because 
the usual nebular diagnostics \cite{oste06} don't work for 
the Weigelt blobs.  For example, [O~III] \lam 4363 and \lam 5007 
are too weak because of $\eta$ Car's low oxygen abundance, 
and [N~II] \lam 5755/\lam 6583 is too sensitive to density in this 
environment.

Using a few [Fe~II] lines measured in blobs B+C+D with the pre-1997 
HST/FOS instrument, and assuming that $T_e \sim 8000$ K, Davidson 
et al.\ estimated $E_{B-V} \sim  0.55$ \cite{kd95}.  Figure 9 shows 
results of a more detailed analysis by Hamann et al. \cite{hama99} 
using every reliably measured [Fe~II] line in the visual and red 
spectrum of blobs B+C obtained in March 1998 with HST/STIS.  Here 
each [Fe~II] line's observed flux $F(\mathrm{observed})$ is 
plotted relative to a theoretical value $F(\mathrm{theoretical})$ 
that assumes LTE conditions at temperatures 6000 and 10000 K, 
$E_{B-V}=0.42$, and a standard reddening curve with $A_V/E_{B-V} = 3.1$ 
(atomic data from \cite{joha77feii}).  Since $\eta$ Car has an 
abnormal $A_V/E_{B-V}$ ratio as noted above, the $A_V$ values in 
Figure 9 should be regarded as estimates of the quantity $3.1 E_{B-V}$, 
smaller than the true $A_V$.    

  \begin{figure}[ht]    
  \center
  \includegraphics[scale=0.4]{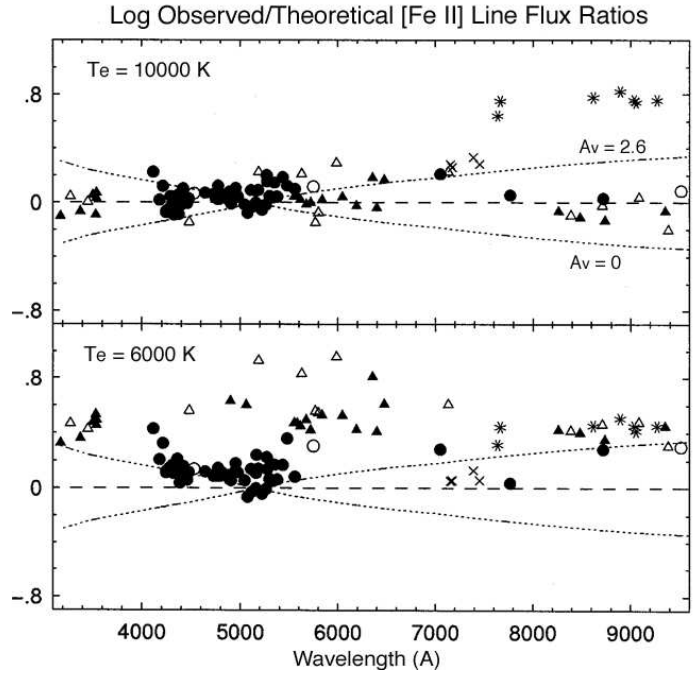}
  \caption{Observed [Fe~II] line fluxes compared to LTE predictions at 
    assumed temperatures 10000 K (top panel) and 6000 K (bottom), if 
    $E(B-V) = 0.42$ ($A_V = 1.3$ with a normal reddening law).  The 
    vertical scale is $C + \log_{10} F_\mathrm{obs} / F_\mathrm{theor}$,  
    normalized to be zero for [Fe~II] $\lambda$5159.  Lines  
    show expected values for $E(B-V) =$ 0, 0.42, and 0.84 mag, 
    with a standard reddening curve.  Each point represents a different 
    [Fe~II] line in the spectrum of Weigelt blobs B+C.  Symbols indicate 
    approximate upper-state energies:  asterisks for $E_{up}< 1.7$ eV,  
    crosses for 1.7 to 2.5 eV,  circles for 2.5 to 3.7 eV,  and triangles 
    for $E_{up} > 3.7$ eV. } 
  \end{figure}

No temperature-and-reddening combination fits all the data.    
Levels above 2.5 eV (circles and triangles in Fig.\ 9) agree fairly well 
with $T \sim 10000$ K and $E_{B-V} \sim 0.4$, but this choice  
under-predicts lower-excitation lines (asterisks).  The latter are more 
consistent with $T \sim 6000$ K and $E_{B-V} \sim 0.8$.  Qualitatively a 
discrepancy like this can occur merely because  there is a range of 
temperatures, so the high-excitation lines preferentially represent the 
highest $T$ -- especially if dust within each blob causes 
$E_{B-V} \sim$ 0.1 or 0.2 mag of {\it local\/} reddening. Quantitatively, 
though, this explanation requires a temperature range broader than 
6000--10000 K.

Most likely the actual gas temperatures are near 6000 K based on 
the lowest energy levels (Figure 9), while levels above 
2 eV are over-populated by continuum pumping. Permitted Fe~II transitions     
absorb UV star light and populate states above 4 eV, followed by 
a cascade through the lower states.  This ``continuum fluorescence'' 
mechanism affects Fe~II emission in AGN broad-line regions 
\cite{netz83,will85}, LBV winds \cite{hill01,hill98},   
H~II regions \cite{vern00}, and probably the ``strontium filament'' 
region in the Homunculus (\cite{bau02,vern02} and \S5 below).
Eta Car has long been recognized as a good locale for fluorescence 
in general (e.g., \cite{kd71,kd95,HDJ94,viot99}),  and the likely 
role of Fe~II continuum fluorescence in the Weigelt blobs became 
clear in the late 1990s \cite{hama99}.

[Ni~II] lines provide additional contraints. For example, 
the strong $\lambda\lambda$7378,7412 lines have an upper level about 
1.8 eV above the ground state, while $\lambda\lambda$7255,7308 
share an upper state near 2.9 eV.  Their ratios are nearly immune 
to reddening but sensitive 
to temperature.  In the high density limit with LTE populations, the 
theoretical $\lambda$7378/$\lambda$7308 ratio is about 21 for 
$T \approx 5000$ K and 5 for $T \approx 10000$ K \cite{hama94}.  
The observed value in blob D at 2002.5 was about 6, broadly consistent 
with the [Fe~II] results if there is some non-LTE over-population 
of the Ni$^+$ upper states.  

During the 2003.5 event, these line ratios signaled a drop in either 
the  temperature or the amount of fluorescence excitation. 
In blob D, for example (\S3), [Fe~II] 
and [Ni~II] lines with higher energy states tended to weaken more 
than the lower-excitation lines.  While this fact might have 
something to do with declining amounts of continuum photo-excitation, 
the [S~II] changes described in \S3 are more readily interpreted 
as a temperature decrease of ${\Delta}T \sim -700$ to $-900$ K. 

Thus the observed changes in the line ratios suggest a temperature 
decline over 2--3 months. This is much longer than the radiative 
cooling time. For the nominal composition and physical conditions we 
derive for the blobs (see below), a representative cooling 
time is 1 to 10 days based on calculations in 
\cite{dalg72} and G.\ J.\ Ferland's unpublished {\it Hazy\/} 
ionization-code manual. In addition, the blobs are located just several 
light-days from the star.  Thus it appears that the temperature responded  
to changes in the spectral energy output from the central source, with only 
small delays due to cooling and light travel times (see also \S5).

\subsection{Densities}     

The rich emission line spectra provide various density indicators.  
The most reliable involve ratios of lines 
with similar upper-state energies, i.e., members of a multiplet, to 
minimize the temperature sensitivity.  
Such ratios can be sensitive to electron densities $n_e$ if they are 
within an order of magnitude or so of the critical densities for 
collisional de-excitation \cite{oste06}.  
Most estimates for the Weigelt blobs provide only lower limits to 
$n_e$ because each observed line ratio is near the high-density 
limit.   Hamann et al.\ \cite{hama99} found these results:   
$n_e > 10^4$ \cmn\ based on [S~II] \lam 6716/\lam 6731, 
$n_e > 10^6$ \cmn\ from [S~II] \lam 4069/\lam 6731, 
$n_e > 10^7$ \cmn\ from [Fe~II] \lam 7155/\lam 8617, and 
$n_e \sim 10^8$ \cmn\ to $10^9$ \cmn\ from [Ni~II] \lam 3439/\lam 3993 
and \lam 7412/\lam 7387.  Wallerstein et al. \cite{wall01} estimated 
$n_e \geq 10^7$ \cmn\ based on [S~II] \lam 4068/\lam 4076.  
The classic ratio [O~II] \lam 3729/\lam 3726 is not 
detected due to the low oxygen abundance, and the analogous 
[N~I] \lam 5201/\lam 5198 lines are severely blended with Fe~II 
and [Fe~II].  Later work \cite{vern02}  produced 
similar results using photoionization models.

The spatial and spectral complexity of the inner ejecta suggest  
that there is a range of densities.   The estimates quoted above apply to 
low-ionization gas and they are skewed toward high densities which 
give the largest emissivities.  There are no simple line-ratio 
density indicators for the more highly ionized gas, but we obtain some 
constraints from the observed time scales for H~I and He~I emission 
changes during spectroscopic events.  Such changes cannot be much faster 
than the recombination time $t_{rec} \sim 1/{\alpha} n_e \,$, 
where $\alpha$ is a recombination rate coefficient. At the 
beginning of the 2003.5 event, the H~I and He~I line fluxes dropped 
substantially in less than 10 days \cite{hart05,dam08a,dam08b}. 
This fact implies densities $n_e > 10^6$ \cmn\ and $n_e \geq  10^7$ \cmn\ 
in the He~I and H~I emitting regions, respectively. 

We conclude that $n_e \sim 10^7$ -- $10^8$ \cmn\ is a reasonable 
estimate for the main emitting regions in the blobs, but 
other density regimes may also be present.

\subsection{Composition}    

The composition of the ejecta is relevant to dust formation in the 
stellar wind and to nucleosynthesis and mixing processes in the stellar 
interior. The amount of CNO processing is of particular interest for $\eta$ Car. 
Hydrogen burning via the CNO cycle produces no net change in the 
total number of C + N + O nuclei, but the reaction rates in   
equilibrium lead to a net conversion of C and O into N. Spectra  
of the ``outer ejecta'' (just outside the nominal outer boundary  
of the bipolar Homunculus) show that nitrogen exceeds 
C + O there  \cite{kd86}.  Inside the Homunculus  
the logarithmic N/O abundance relative to 
solar is very large, $2.0\leq {\rm [N/O]}\leq 2.5$ \cite{dufo97},
while the ejecta far outside it have nearly solar N/O \cite{sm04,smit05}.
The outer ejecta thus appear to contain the first CNO processed gas 
to be expelled, possibly just before the $\sim$1843 eruption 
(see also \cite{weis99}). The stellar wind today is also 
CNO-processed \cite{hill01}.

If the blobs and inner ejecta were expelled after the Homunculus
(\S1), then they too should contain heavily CNO processed gas. 
The most reliable abundance estimates rely on 
lines that form in the same physical conditions. Hamann et al.\ 
\cite{hama99} used the N~III] $\lambda$1750 and O~III] $\lambda$1664
inter-combination lines to estimate [N/O] $>$ 1.8 (i.e., 
$n(\mathrm{N})/n(\mathrm{O}) > 60$) in blobs B+D. They also 
estimate [Fe/O] $\sim$ 2.0 to 2.3, based on [Fe~II] \lam 8617/[O~I] 
\lam 6300 and [Fe~II] \lam 7155/[O~I] \lam 6300, where the factor-of-two 
uncertainty comes mainly from the uncertain density. These results 
are consistent with an estimate that [C/Ne] and [O/Ne] are both roughly 
$-$1.7 to $-$2.0 based on photoionization models \cite{vern05}.  
We conclude that the blobs and associated high-ionization gas are 
heavily CNO-processed.

The dust content of the blobs is more uncertain. The fact that we see a 
reflected spectrum of the star in the blobs indicates that dust 
is present.  Mid-IR observations show warm dust in the inner ejecta 
\cite{ches05, s+03ir}. The strutures seen in the mid-IR images closely correspond with
the knots seen in the visible although they are  not spatially coincident.
The visible structures,
dominated by scattering, trace the  {\it walls} of the dense clumps of dust, while the
infrared structures are identified with the  emission from hot dust, probably the
external layers of the clumps.   This observational evidence is consistent with the 
theoretical  dust temperatures, which indicate that dust can survive closer to
the star than the nearest Weigelt blob B \cite{s+03ir,kdrh97}. The infrared flux 
appears to be decreasing from 2002 to 2005 \cite{art}, but is not correlated with the 
2003.5 event or with orbital phase. It may be due to enhanced dust destruction in response to the increased stellar flux.

To avoid projection effects and determine how much dust actually 
resides {\it within\/} the line-emitting blobs, we can examine the 
gas phase depletions of refractory elements like Fe, Cr, Ni, Ti and Ca 
compared to non-refractory elements like C, N, O, Ar and S.   In 
cool interstellar gas clouds in our Galaxy, refractory elements are 
typically depleted by factors of 10 to $>$100 because they are locked 
up in dust grains \cite{sava96}.  Iron, in particular, is depleted by 
a factor of $\sim$200 in cool Galactic clouds. In the Weigelt blobs, 
however, one study of [Fe~III] lines found the Fe/H abundance 
to be roughly half solar \cite{vern05} .  If we combine this 
with an estimate by Hillier 
et al. \cite{hill01,HGN06} for solar Fe/H in the star, we conclude 
that iron is not strongly depleted in the vicinity of the blobs.      
This result is consistent with a cursory inspection of the low-ionization  
blob spectra, wherein emission lines of Fe, Ca and Ti are well 
represented compared to the lines of non-refractory elements like S 
and even the grossly overabundant N. Thus the depletions of refractory 
elements are much smaller in the blobs than in cool Galactic clouds.

There are two other hints that the dust-to-gas ratio is small
in the blobs. First, the strong fluorescent emission lines of Fe~II and 
O~I pumped by Ly$\alpha$ and Ly$\beta$, respectively,  require 
many scattering events in the Lyman lines (\S4.5 below).  A ``normal'' 
Galactic 
amount of dust would destroy the Lyman line photons before they 
are absorbed substantially Fe~II or O~I.  Second, energy budget 
considerations indicate that the blob material (like the Homunculus 
on much larger scales) has a relatively unobscured view of the central 
object \cite{kd95}.  If much dust is present, it must have a patchy 
distribution so that starlight can largely avoid it by scattering 
\cite{MDK06,HGN06}.

\subsection{Fe~II and Fluorescent Line Emission}  

Spectra of the blobs and inner ejecta are strongly affected by resonant 
fluorescence.  Fe~II has by far the richest spectrum of known fluorescent 
lines, including a spectacular pair at 2507.6 and 2509.1 \AA\ 
(Figure 5, \cite{viot89,kd95,kd97,joha93,HDJ94}). 
There are also fluorescent lines of O~I, Cr~II, Fe~III, Ni~II and 
possibly Mn~II \cite{joha95,joha00,zeth01cr}. Their upper energy states 
are vastly over-populated because of accidental wavelength coincidences 
with H~I Lyman lines. An interesting exception is Mn~II, which apparently 
absorbs a strong UV line of Si~II \cite{joha95}.  Fe~II and O~I 
fluorescence has been discussed thoroughly \cite{oste06,elit85,gran80,sigu98} 
for a variety of stellar environments, including symbiotic stars and  
of other early-type stars with dense circumstellar envelopes 
\cite{rudy00,carp88,pens83,hart00,joha83,joha84,hama88,hama89}.  
In general this phenomenon can help us diagnose physical and radiative 
conditions in the emitting regions.

The fluorescence in Fe$^+$ is ``pumped" by Ly$\alpha$ and yields a unique 
pattern of emission lines. The primary cascade lines appear in the UV 
between $\sim$1800 \AA\ and $\sim$3000 \AA\ and also in the far-red between 
$\sim$8000 \AA\ and $\sim$10000 \AA\ (\cite{HDJ94} and refs. therein). 
Figures 4 and 5 above show examples of Fe~II fluorescent lines 
in the blob D spectrum. These lines 
would not be detectable without this form of excitation.

The strongest fluorescent lines in $\eta$ Car are the Fe~II 
$\lambda\lambda$2508,2509 shown in Figure 5. Their particular excitation 
has been discussed extensively by Johansson et al.\ \cite{joha93,joha98}.   
They constitute an interesting puzzle because they appear far too strong 
compared to other lines arising from the same upper states; transitions to 
particular lower states are anomalously favored. Johansson et al. \cite{joha98} and 
Johansson \& Letokhov \cite{joha04} proposed that stimulated emission 
is responsible -- a natural UV laser!  This hypothesis is controversial 
because it seems incompatible with simple estimates of the photon densities 
\cite{zeth99,kdrh97};  more work is needed.  In any case, Ly$\alpha$ 
fluorescence clearly does control 
the excitation of $\lambda\lambda$2508,2509.

Figure 4 also shows fluorescent O~I $\lambda$8446, which represents 
a secondary cascade from an energy state pumped by H~I Ly$\beta$. 
The importance of fluorescent excitation in this case can be deduced from 
the relative strengths of the primary cascade lines in the near-IR 
(e.g., $\lambda$11287/$\lambda$13165) and from the strength of $\lambda$8446 
compared to non-fluorescent O~I $\lambda$7773 \cite{HDJ94}.  The O~I 
resonance wavelength differs from from Ly$\beta$ by only 0.04 \AA\ or 
12 \kms .  Given the high Ly$\beta$ opacity expected in these regions,  
Ly$\beta$ photons incident from the outside would have little effect on the 
O$^0$ excitation. Therefore O~I fluorescence must be driven by Ly$\beta$ 
photons created within the blobs;  we will return to this point below.

Simple considerations of the fluorescence processes lead to useful 
constraints on the physical conditions.  For example, the Fe$^+$ transitions 
that absorb Ly$\alpha$ arise from metastable lower states which must be 
significantly populated. These populations can be maintained by 
collisions if the gas densities are above the critical 
values for those levels, $n_e \geq 10^6$ \cmn , consistent with our 
estimates in \S4.3 above. 

Another constraint involves ionization. Strong O~I emission requires 
a significant amount of neutral gas, since $n(\mathrm{O}^0)/n(\mathrm{O}^+)$ 
is closely coupled to $n(\mathrm{H}^0)/n(\mathrm{H}^+)$ by charge exchange 
reactions.  The Fe~II emission regions are also expected to have appreciable 
amounts of H$^0$  (\S5 below and \cite{vern02}). However, Ly$\alpha$ and 
Ly$\beta$ photons must be abundant in order to drive the fluorescence. Any 
Lyman line that is incident from the outside will be blocked by the extremely 
large line opacities of H$^0$.  Therefore, the fluorescence observed 
in Fe~II, O~I, etc. is caused by Lyman line photons that are 
created locally inside the emitting regions (see also below and 
\cite{HDJ94}). This requires a particular ionization balance 
with enough neutrals 
to maintain sufficient O$^0$, Fe$^+$, etc., but also enough H$^+$ 
to produce Lyman line emission. In a photoionized gas, this balance 
is achieved in zones of partial ionization just behind 
(i.e., on the more-neutral side of) the H$^+$--H$^0$ recombination front 
\cite{vern02,bal04}.

One can think of Fe~II fluorescence as an escape route for 
Ly$\alpha$ photons that are otherwise trapped.  The low 
Fe/H $\sim$ Fe$^+$/H$^0$ abundance ratio means that such photons 
will scatter many times from hydrogen atoms before being 
absorbed by Fe$^+$.  Some of this absorption occurs in Fe~II 
lines that have relatively poor wavelength coincidences with 
Ly$\alpha$ (1215.67 \AA ).  For instance, the strong Fe~II $\lambda$2508 
line is pumped by a UV transition  offset by 630 \kms .
Therefore, if the Ly$\alpha$ line profile in this gas is symmetric, it cannot be 
much less than 1260 \kms wide.  For another strong fluorescent line, 
Fe~II \lam 9123, the corresponding value is 1340 \kms .  
Since fluorescent Fe~II lines with poorer wavelength coincidences 
are absent, 1300 \kms\ is a fair estimate of the full width of the 
exciting Ly$\alpha$ line within the gas \cite{HDJ94,hart05}. Similar 
results have been derived from the Cr~II fluorescence features 
\cite{zeth01cr}.

Locally-emitted Ly$\alpha$ photons can indeed be distributed across this 
wide range in apparent velocities because of natural broadening.  If the 
line-center optical depth is large,  $\tau_0 \geq 10^4$, then the 
Ly$\alpha$ absorption profile is dominated by damping wings and 
FWHM $\sim 0.18\, v_D\,\tau_0^{1/3}$, where $v_D$ is a characteristic 
doppler width \cite{elit86}. Our estimate FWHM(Ly$\alpha$) $\sim 1300$ \kms\ 
implies that Ly$\alpha$ has $\tau_0 \sim 3\times 10^8$, which corresponds 
to an H$^0$ column density of $N({\rm H}^0)\sim 3 \times 10^{21}$ \cmN\ 
if the velocity distribution is thermal and $T_e\sim 7000$ K. 
This estimate agrees well with theoretical predictions for dense 
Fe~II emitting regions \cite{bal04,elit85}. If the Fe~II emitting region 
in $\eta$ Car is $\sim$ 20\% ionized \cite{vern02}, then the total 
hydrogen column would be roughly $N({\rm H}) \sim 4\times 10^{21}$ \cmN . 
This is much smaller than the probable column density through an entire
Weigelt blob (\S1), which is consistent with the fluorescent lines 
forming in boundary layers between the neutral and ionized gas.

Another important constraint comes from the O~I fluorescence. In 
low-density H~II regions where the optical depth in H$\alpha$ is not 
large, Ly$\beta$ is converted, after only about 10 scattering 
events, into H$\alpha$ plus Ly$\alpha$ photons which then escape:  
``Case B recombination.''  This situation cannot produce O~I fluorescence  
because a typical Ly$\beta$ photon is scattered by H$^0$ about 
$5\times 10^4$ as often as by O$^0$, leaving little opportunity for 
O$^0$ to absorb Ly$\beta$ \cite{kdhn79,gran80}.   O~I fluorescence 
therefore requires large optical depths in H$\alpha$ in order to trap 
the H$\alpha$ photons and inhibit the conversion of Ly$\beta$ into 
H$\alpha$ + Ly$\alpha$.  This in turn requires a large population  
in the $n=2$ level of H$^0$.  If we could view this gas against the 
background of a bright continuum source, we should see 
strong narrow absorption in the Balmer lines (see \S4.6 below)!

Finally, it has been suggested that the Ly$\alpha$ photons needed for 
the Fe~II fluorescence come from the stellar wind 
rather than in situ emission in or near the Weigelt blobs 
\cite{kd01b,vern02,hart05,joha04}.  This might seem plausible because 
i) the Ly$\alpha$ intensity from the star should be much stronger 
than the adjacent stellar continuum \cite{hill01},  and 
ii) the width of the Ly$\alpha$ profile can be FWHM $\sim 1340$ 
\kms\ because the wind speeds can exceed 500 \kms\ \cite{hill01,s+03lat}. 
However, it is not clear that the wind's Ly$\alpha$ profile 
really is this broad; the observed stellar Balmer lines are narrower 
(Fig. 7, \cite{HGN06,kd05}.  Several other stars with strong nebular 
Fe~II fluorescence have even narrower H~I lines than $\eta$ Car, 
though their fluorescent spectra also require Ly$\alpha$ with 
FWHM $\sim 1300$ \kms\ \cite{hama88,hama89}.  Therefore an external 
source of broad H~I emission lines does not seem to be important
for the fluorescent Fe$^+$ excitation.  A more serious concern is that the  
large Lyman line opacities would prevent external photons from 
penetrating the blobs to drive the fluorescence. We noted above
that Ly$\beta$ photons from the outside cannot play any role in the 
O~I pumping. The large Ly$\alpha$ opacities in the blobs imply that 
external Ly$\alpha$ photons could pump Fe~II only  
in transitions that are far removed from the Ly$\alpha$ line center;  
the other Fe~II lines need to be pumped by locally created Ly$\alpha$. 
Thus there would need to be two fluorescent 
processes operating at the same time and varying 
in unison during the spectroscopic events. Moreover, it is not 
obvious why the star's Ly$\alpha$ intensity would vary as needed because,
for example, the stellar Balmer line fluxes change by only a factor of 
$\sim$2. Overall, the observed Fe~II fluorescent 
lines are more easily understood if they are linked to the 
ionization and the local creation of Ly$\alpha$ photons. 

\subsection{Narrow Nebular Absorption Lines}  

Narrow absorption lines of H~I and some low-ionization metals appear 
across the 1--2\arcsec\ core of the Homunculus, including the 
central star and the Weigelt blobs 
\cite{kd99,hama99,stis99,gull01,gull06,joha05,niel07}. 
Figure 7 shows, for example, narrow absorption in H$\delta$ and H$\gamma$ 
in the spectrum of blob D.  These features are clearly not related to 
the broad P Cygni wind profiles.  At least some of them form in the 
inner ejecta (see below), while others apparently arise farther out 
in the ``Little Homunculus" or in the outer shell surrounding the 
Homunculus itself \cite{niel07,gull06,gull01}.  This situation is highly 
unusual; one does not generally see 
Balmer absorption in the ISM or even in denser-than-average nebulae 
because very few H~I atoms there are in the $n = 2$ level.

The metal absorption lines have low ionizations typified 
by Fe~II and Ti~II. In spectra towards the star they show at least 
30 distinct velocity components, with the strongest features near  
$-$146 and $-$513 \kms .  In the blob spectra the absorption components 
are less distinct and their velocities are different \cite{niel07}. 
A detailed analysis of the stellar spectrum \cite{gull06,niel07} 
shows that many of the Fe~II and Ti~II lines arise from metastable 
excited states, so densities in the absorbing gas must be near or above 
those states' critical values.  Gull et al.\cite{gull06} estimate that the strong 
system at $-$146 \kms\ has $n_e \sim 10^7$ to $10^8$ \cmn  and 
$T \sim 5700$ to 7300 K, located roughly 1300 AU  from the star.  
The relevant column density in Fe$^+$ is $\sim \, 5\times 10^{15}$ \cmN . 
If we assume that Fe/H is solar and all of the iron is singly ionized,
then the corresponding hydrogen column is $N_H \sim 2 \times 10^{20}$ \cmN . 
They  also note that metastable Fe$^+$ levels below $\sim$ 3.2 eV 
are approximately in LTE, while those at higher energies are 
overpopulated compared to LTE -- similar in this respect to the 
Weigelt blobs as discussed above.  The estimated location places the 
absorbing gas within the inner ejecta, at $r < 1$\arcsec , 
but it is clearly distinct from the blobs since it has a different 
velocity and lies along our line of sight to the central star.

Balmer absorption lines in a nebular environment are surprising 
because they require significant 
populations in the $n=2$ level of H$^0$.  A line-center optical depth 
of $\tau_0 \geq 1$ in H$\gamma$, for example, implies a column 
density $N(n=2) \geq 3 \times 10^{13}$~\cmN\ if the doppler velocities 
are thermal with $T \approx 7000$ K. This requires a dense gas that 
is neutral enough to have H$^0$ but also ionized enough to populate the $n=2$ 
level.  This situation is believed to occur in the broad emission line regions 
of quasars \cite{ferl79,kdhn79}, and Balmer line absorption has been 
observed directly in quasar outflows where the densities and ionizations 
might be similar to the inner ejecta of $\eta$ Car \cite{hutc02,hall07}.

One way to populate the H~I $n = 2$ level is by collisions in a warm 
gas where this level is thermalized, such that that the 
downward rate due to electron collisions exceeds the  net rate 
for radiative decays.  At first sight this would require an absurdly  
large electron density -- except that almost every radiative decay is 
nullified, macroscopically speaking, when the fresh Ly$\alpha$ 
photon is immediately absorbed in an excitation event.  Therefore 
our {\it net\/} radiative decay rate includes only the few Ly$\alpha$ 
photons that escape from the vicinity.  In this case the minimum density 
for strong collisional de-excitation is roughly given by 
  \begin{equation}
  { {n_e \, q_{21}}\over{A_{21} \beta}} \approx 
     {{n_e \, \tau_0}\over{n_{cr}} } \geq 1 \, ,   
  \end{equation}
where $q_{21}$ is the downward collision rate coefficient, 
$\tau_0$ is the line-center optical depth in Ly$\alpha$, 
$\beta \sim 1/\tau_0$ is the escape probability for Ly$\alpha$ photons,
and $n_{cr} \approx A_{21}/q_{21} \sim 10^{17}$~\cmn\ is the critical density 
for the $n = 2$ level at $T_e \sim 7000$K in the absence of Ly$\alpha$ 
entrapment.  For $n_e \leq 10^9$~\cmn , thermalization requires 
$\tau_0 \geq 10^8$ and thus a total H$^0$ column density 
$N(n=1) \sim N({\rm H}^0) \geq 10^{21}$~\cmN\ , assuming a thermal 
velocity dispersion.   This corresponds to 
$N(n=2) \geq 2\times 10^{14}$ \cmN\ in LTE at 7000 K, easily 
sufficient to produce $\tau_0 > 1$ for the Balmer absorption lines.

These physical conditions are reasonable for the partially ionized gas 
associated with the Weigelt blobs. In fact, they resemble what 
we inferred from the fluorescent emission lines (\S4.5).  However, 
thermalization is an extreme requirement. Balmer absorption 
lines can occur at values of $N({\rm H}^o)$ and $\tau_o({\rm Ly}\alpha )$, 
below the thermalization limit if recombination is also important 
for creating Ly$\alpha$ photons that are subsequently trapped. 
The inability of these Ly$\alpha$ photons to escape can lead to 
$n=2$ populations that are significantly enhanced 
relative to LTE \cite{hall07}.

This general scheme for observable Balmer absorption is supported by 
measurements of damped Ly$\alpha$ and Ly$\beta$ absorption lines in 
spectra of the  central star \cite{HGN06}. The origin of the damped 
lines is uncertain 
because they are too broad to measure their kinematics. However, they 
clearly form in the nebular environment of $\eta$ Car and a likely 
location is in the Balmer line absorber discussed above. 
The neutral hydrogen column density derived for the 
damped absorber, $N({\rm H}^o)\sim 3\times 10^{22}$~\cmN , is 
also consistent with the conditions needed for Balmer line absorption.

Another factor that might play a role is the metastable nature of 
the H$^0 \; 2s$ state \cite{joha05}.  Radiative decays from $2s$ 
occur primarily by  2-photon emission with transition probability 
$\sim$8~s$^{-1}$, compared to $\sim$$6\times 10^8$~s$^{-1}$ 
for Ly$\alpha$ decay from $2p$.  Therefore the $2s$ level thermalizes 
at much lower density than $2p$.  
On the other hand, collisional mixing between the $l$ states 
(e.g.\ $2s \leftrightarrow  2p$) may help to depopulate $2s$.
Detailed calculations are needed to examine the various processes 
controlling H$^0$ ionization and $n=2$ population in environments 
consistent with the absorbers in $\eta$ Car\footnote{
   Johansson et al. \cite{joha05} argued that the $2s$ population 
   is regulated by absorption of Ly$\beta$ photons from the central 
   object.  However, that scheme ignores the large Lyman line opacities 
   in the absorbing nebula (see also \S4.5). We also note that 
   most of the environments favored by their calculations for Balmer 
   line absorption would be optically thick to Thomson scattering 
   at all wavelengths.}.  

In any case, the narrow absorption lines clearly represent important gas 
components in the Homunculus. Some of the absorbers reside in the 
inner ejecta with large column densities and physical conditions 
similar to the Weigelt blobs. They appear to be blob-like material 
that happens to be viewed against a bright background of direct 
or reflected starlight. The narrow Balmer absorption lines seen 
toward the blobs might actually form in an outer ionized layer 
of the blobs themselves (\S4.1). However, the distributed appearance 
of the Balmer and other narrow absorption lines across the inner 
ejecta, with a range of velocities, shows that these absorbing 
regions are much more extended than the individual Weigelt blobs. 
They might also contain a significant amount of mass (\S6 below).

\subsection{Mass of the Weigelt Blobs and Inner Ejecta}

The mass of ejected material has direct implications for the nature,   
evolution, and instabilities of the central star.  We can estimate 
this mass from the strengths of forbidden emission lines.  This 
is fairly straightforward because the relevant level populations are 
close to LTE (see above);  the observed flux is therefore proportional 
to mass rather than mass times density.   Davidson et al. \cite{kd95} 
used the [Fe~II] \lam 5376 flux measured in {\it HST} observations 
of a 0.3{\arcsec} region including Weigelt blobs B+C+D, and 
found $\sim$$0.002$ M$_{\odot}$ assuming all of the iron is 
Fe$^+$, solar Fe/H abundances, 3 magnitudes of 
extinction, and LTE at $T_e=8000$ K. This estimate is surely a lower 
limit because it ignores possible depletion of iron into dust grains 
and it applies only to the emitting regions of [Fe~II] \lam 5376.
(It excludes the more highly ionized zones, see \S5).  Nonetheless, 
if we adopt a diameter 0.1\arcsec  $\approx$ 230 AU 
for each blob (\S1), we find that this mass corresponds to an average 
density $n_H \sim 3\times 10^7$ \cmn  within them, reasonably consistent 
with our estimates above (\S4.3).

We can estimate the mass outside the blobs by using emission lines 
measured through a larger 1\arcsec  aperture.  Starting with the 
measured flux in 
[Fe~II] \lam 7155 \cite{HDJ94}, we reduce this flux by 20\% 
to represent only the narrow (non-stellar) 
emission component, correct for 2 magnitudes 
of red extinction, assume solar Fe/H, and LTE populations with 
$T_e=7000$ K.  The result, $\sim$$0.006$ M$_{\odot}$, 
is three times larger than the Davidson et al.\ estimate for 
the inner 0.3\arcsec.   A similar calculation applied to 
[Ar III] \lam 7136 in the same ground-based data (but including its 
entire flux because it appears to be entirely nebular, 
\S4.1), yields $\sim$0.002 M$_{\odot}$ for higher ionization gas.

The narrow absorption lines might provide a rough estimate of the nebular 
mass independent of extinction.  For instance, if the Balmer line 
absorber covers an area of $1^{\prime\prime} \times 1^{\prime\prime}
\approx  2300^2$ AU$^2$ and its average column density is 
$N(H^0)\sim 5\times 10^{21}$ \cmN\ (well below the value measured 
toward the star but above the minimum needed for thermalization, 
\S4.5), then the total mass in this absorber is 
$\sim$$0.005/f_0$ M$_{\odot}$, where $f_0 = H^0/(H^0 + H^+) < 1$ is 
the neutral fraction.

To some degree, we can simply add these mass estimates together because 
they represent different gas components.  Doing this we find a minimum 
total mass of $\sim$0.013 M$_{\odot}$ within the central 
$1^{\prime\prime} \times 1^{\prime\prime}$. However, the values based 
on emission line fluxes are only lower limits because i) they 
probe just the optimal emitting regions for particular lines, and 
ii) the extinction corrections may be larger if, as expected, 
the obscuration is substantially gray (\S4.2) or the dust 
distribution is patchy.  (Fainter regions may be those with more 
extinction rather than less emission, \S1). Based on these 
considerations, the total mass in the {\it inner\/} ejecta is 
most likely in the range $\sim$0.02 to 0.05 M$_{\odot}$ 
(see also \cite{kd97}).

\subsection{The Strontium Filament}

The ``strontium filament'' is a patch of nebulosity located several 
arcsec northwest of the star.  It is remarkable for its 
Sr II and [Sr II] emission lines as well as some extraordinarily 
low-ionization features \cite{zeth01sr,zeth01cr,bau02,hart04,bau06}. 
Not really a filament, this structure is much larger than the 
Weigelt blobs and has more complicated kinematics. Studies of the 
Sr filament may be helpful for understanding the inner ejecta.

Its spectrum is essentially a lower-ionization version of 
that emitted by the Weigelt blobs.  It is dominated by emission 
from species such as C$^0$, Mg$^0$, Ca$^0$, Ca$^+$, Sc$^+$, 
Ti$^+$, V$^+$ and Mn$^+$, in addition to the signature Sr$^+$ 
lines \cite{hart04}.  Its iron spectrum has more Fe~I and [Fe~I] 
than Fe~II and [Fe~II], and there are none of the fluorescent 
lines of Fe~II, O~I, etc., that require partially ionized gas 
(\S4.5). There are also no H~I or He~I emission lines;  the 
hydrogen must be essentially neutral.

Spectroscopically, the Weigelt blobs most resemble the Sr filament 
during 
a spectroscopic event.  At such a time the blobs' H~I and He~I lines 
become extremely weak, fluorescent lines disappear, and the lowest 
ionization features like [Ca~II] and Ti~II and [Ti~II] strengthen (\S3). 
The blobs never become as neutral as the Sr filament, but overall the 
conditions then appear to be similar.  Calculations of multi-level 
Sr$^+$ and Ti$^+$ atoms in the Sr region \cite{bau02,bau06} suggest 
that the free electron densities there are of order $n_e\sim 10^7$ \cmn\ 
at temperatures of $T_e\sim 6000$ K  (cf.\ \S4.2 and \S4.3). 
Considering that the gas is mostly neutral, the total hydrogen density 
may be substantially higher.  The energy source for this region's 
line emission is believed to be incident stellar radiation.  However, 
the lower ionization in the filament cannot be explained merely by 
its distance from the star.  The stellar spectrum seen by the filament 
appears to cut off sharply above $\sim$8 eV, limiting the ionization 
state to neutrals and some singly-ionized species in the iron group 
\cite{hart04}. The Weigelt blobs see a harder stellar spectrum, 
even during a spectroscopic event.

The strong metal line emissions from both the blobs and the Sr filament 
are believed to be excited by a combination of collisions and 
photo-absorption of the stellar visible and near-UV flux, i.e., 
by continuum pumping (see below).

\section{Line Formation Physics}  

     The rich emission-line spectra discussed above present many diagnostic 
     opportunities but also a basic problem: How are they produced?   The 
     importance of radiation from the star is evident in the spectroscopic 
     events. For example, the stellar wind P Cygni features reflected by 
     dust in the blobs change at roughly the same time as the in situ 
     narrow emission lines (\S3 and Fig.\ 7).  Changes in the star and 
     in the blobs track each other within a month or less
     (see also \cite{stah05,HGN06,kd05,dam08a}. This fact probably 
     indicates that the blob spectra are responding to changes in   
     radiation from the star (or rather the two stars).  Evidently this 
     is the energy source for ionization and excitation inside the blobs.  
     The other possibility, kinetic energy in the stellar wind, is less  
     powerful and is ruled out by the event timings (also \S3).

 Verner et al. \cite{vern02,vern05} and Mehner et al. \cite{meh10a} 
 used photoionization codes to show that most properties of the blob spectra 
 can, indeed, be matched by dense clouds irradiated by sources like those 
 expected for the primary star and its putative hot companion.   In 
 these models the high-ionization lines of [Ne~III], [Ar~III], He~I, etc., 
 form in (H$^+$) layers directly exposed to the central source, while 
 Fe~II, [Fe~II], Ni~II, [Ni~II], [Ca~II], Ti~II, [Ti II], etc., form in 
 a warm, partially ionized environment behind the H$^0$--H$^+$ 
 recombination front. The predicted temperature there is roughly
 5000--7000 K and the H$^+$/H fraction ranges from 
 $\sim$ 50\% in the warmest regions to $\sim$15\% farther behind the 
 front \cite{vern02}.  Collisions and UV continuum pumping together 
 produce strong emission from Fe~II, [Fe~II] and similar ions.

 Circumstances like these have often been discussed for the Fe~II 
 emission regions of active galactic nuclei \cite{ferl79,netz83,will85,
 vern99,bal04}.  Continuum pumping dominates the excitation of Fe$^+$ 
 and similar ions for energy levels above a few eV \cite{vern02,vern05,
 bau02,bau06,hill98}.  Thermal collisional excitation populates 
 metastable lower states, which serve as launching pads for continuum 
 pumping to the higher states \cite{vern02,bal04}.

 The well-studied line absorber at $-$146 \kms\ appears to be blob-like 
 material seen against the background of the stellar continuum (\S4.6). 
 Specific calculations for that environment 
 indicate, again,  that low states of Fe$^+$ are populated by collisions 
 at $T_e\sim 6400$ K and $n_e \sim 10^7$ to $10^8$ \cmn . UV absorption 
 lines directly measure continuum pumping out of these states \cite{gull05}. 
 Narrow Balmer absorption lines at this same velocity in the stellar 
 spectrum indicate that the low-ionization absorber is accompanied by 
 enough hydrogen ionizations to populate the $n=2$ level of H$^0$.     

 These results  take us a long way toward understanding the spectra 
 of the Sr filament and inner ejecta, but there are complications. 
 The most basic is that the local ionization waxes and wanes 
 with the 5.54 yr event cycle. To some degree we can think of this 
 as a spatial movement of the H$^0$--H$^+$ recombination front.  Between 
 spectroscopic events, such a front must exist somewhere in the region 
 of interest, because we see both low and high ionization lines.\footnote{  
 H$^0$ and H$^+$ can be diffusely mixed in comparable amounts, 
 with no well-defined ionization front, only if the ``photoionization 
 parameter'' $U_\mathrm{H}$ is very small \cite{kdhn79}.  
 Straightforward models do not allow such small values in $\eta$ Car's  
 inner ejecta   \cite{meh10a}.  }   
 During each spectroscopic event, however, the spectral signatures of 
 the (H$^+$) zone dramatically weaken throughout the inner ejecta. The 
 recombination front must then be much closer the central star.

  A key to understanding these ionization changes is in the detailed 
  timing of the transition from spectroscopic high to low states.  
  First one sees a decline in the [Ne~III], [Ar~III] and He~I 
  lines, followed in order by [Fe~III], Si~III], N~II, [N~II] and H~I, 
  according to their ionization energies (\S3, also \cite{dam08a} 
  and refs.\ therein).    Before the event there must be a significant 
  flux with $h \nu > 41$ eV to maintain the Ne$^{++}$. But then, over 
  about 2 weeks, the source cutoff energy slides down to the Lyman limit 
  at 13.6 eV.  This behavior reverses during the recovery phase, but 
  over a longer time.  Meanwhile, most low-ionization features, 
 notably [Fe~II] and the non-fluorescent Fe~II 
 lines, remain fairly steady.  Their ionization and excitation must be 
 less affected by changes in the flux above the Lyman limit. 
 The lowest-ionization lines, [Ca~II] and [Ti II], strengthen during 
 an event -- most likely due to a general shift in ionization from 
 Ca$^{+2}$ and Ti$^{+2}$ toward Ca$^+$ and Ti$^+$.  

 Mehner et al.\ \cite{meh10a}  reported the [Ne~III] and [Fe~III] 
 behavior shown in Figure 10, including a strong peak 
 several months before the 2003.5 event (see \S4.1 above).  These 
 {\it HST/STIS\/} data refer specifically to gas along our line of 
 sight to the star, with a velocity of $-40$ km s$^{-1}$ like the 
 Weigelt blobs. The blobs vary in a similar way, but the peak just 
 mentioned  was not reported in ground-based data \cite{dam08a}.

\begin{figure}[ht]    
\center
\includegraphics[scale=0.5]{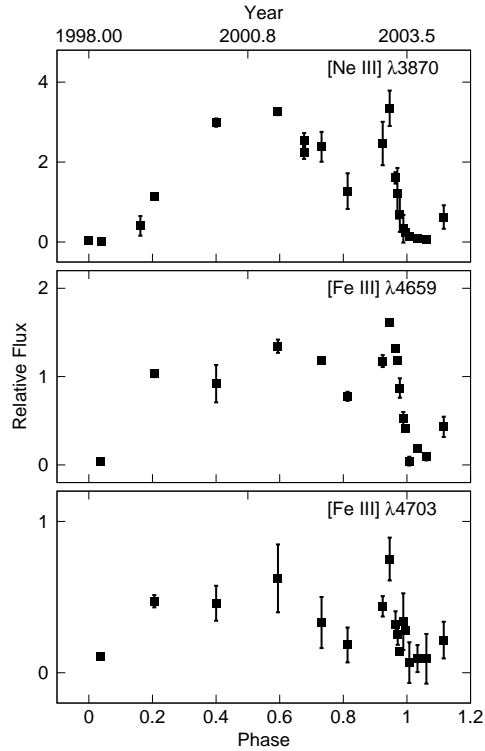}
\caption{The strengths of the narrow [Ne~III] and [Fe~III] emission lines 
  measured in {\it HST/STIS} spectra throughout its spectroscopic cycle. 
  The lines in the blobs vary in a similar way. From \cite{meh10a}. }
\end{figure}

 The Sr filament, on the other hand, has a lower ionization spectrum and 
 remains that way throughout the event cycle \cite{hart04}. 
 Evidently the H$^0$--H$^+$ ionization front never reaches that 
 region, although the stellar flux would normally be strong enough 
 to do so.    Some intervening gas must block the stellar UV continuum 
and impose a cutoff above $\sim$ 8 eV at that location of the filament 
 \cite{hart04}.  The shield must be substantially non-ionized, with large 
  bound-free opacities not just in H~I but also in complex atoms with 
 ionization thresholds near 8 eV.  The leading candidates are Mg$^0$, 
 Si$^0$, and Fe$^0$ with ionization energies 7.6, 8.1, and 7.9 eV. 
  The nature of this UV shielding medium is almost as intriguing 
  as the Sr filament itself, but we should keep in mind that 
  UV extinction may occur throughout a range of locations.

The most puzzling aspect of line emission from the Sr Filament and 
the Weigelt blobs during an event is that neither region appears to 
contain a significant amount of H$^+$.  Hydrogen recombination lines are 
very weak or absent ({\S}{\S}3,4.7).  The photoionization calculations 
 cited above generically predict that the low-ionization emitting zones 
 should be partially ionized.  Some amount of ionization is 
 necessary to provide photoelectric heating and free electrons.  
 If hydrogen and helium are entirely neutral, then the burden 
 for these tasks falls entirely upon the heavy elements and possibly 
 dust grains. This seems problematic because it would lead to 
 very small electron fractions and temperatures 
  insufficient to collisionally excite even the low-energy forbidden 
 lines in species like Fe$^+$ or Fe$^0$. There should be some amount 
 of ionization leading to H~I line emission during all phases 
 of the event cycle. 

The solution to this problem is not obvious.  One possibility  
is that significant H$^+$ exists in these regions 
but has escaped detection.  Quantitative upper limits to Balmer lines
in the Sr Filament have not been reported;  Hartman et al.\ \cite{hart04} 
imply that they are weaker than other lines.  Weak nebular H~I lines  
are difficult to measure near $\eta$ Car because they are blended 
with the dust-reflected broad stellar wind lines.  Blob D as described 
in \S3 appears to have narrow H$\alpha$ and H$\beta$ emission during 
the low-ionization states, but these are weaker than before the event  
and H$\gamma$ and H$\delta$ are hard to detect (Fig.\ 7).  If we 
compare the weak-state narrow H$\beta$ flux to a typical [Fe~II] line, 
we find that the ionized fraction H$^+$/H in blob D during the 2003.5 
event was less than 1/10 of the values 15\% to 50\% predicted by models 
of the Fe$^+$ emitting region \cite{vern02}.\footnote{    
      Here we assume Case B recombination for H$\beta$ and LTE for [Fe~II], 
         with $n_\mathrm{H} \sim 10^7$ \cmn\ and $T \approx 7000$ K. }   
	 The disappearance of Fe~II fluorescence pumped by in situ Ly$\alpha$ 
	 (\S4.5) also suggests that the Fe~II zone is effectively H$^0$ during 
	 an event.  Meanwhile the temperature drops by only about 10\% and 
	 collisionally excited [Fe~II] lines remain nearly steady, as though 
	 they have nothing to do with the changing ionization.

Another possibility is that shocks or turbulence in the outflow 
provide just the right amount of heat and free electrons without 
significantly ionizing hydrogen.  This process might be supplemented 
by absorption of the stellar flux by heavy elements and possibly 
dust grains.   It is not at all clear whether this scenario is viable, 
but it is worth investigating because these processes are unrelated 
to the incident far-UV flux and the hydrogen ionization.  They 
might provide a natural explanation for the steadiness of some 
low-ionization emission throughout the 5.54 yr event cycle.

A third possibility is that the gas is too cool for collisional 
excitation  but some other process drives the low-ionization 
emission. Continuum pumping is expected to play an important role, 
which may be enhanced if non-thermal motions (e.g., turbulence)
broaden the lines and thus enhance the photo-excitation rates 
\cite{netz83,vern02,bal04}.  Continuum pumping ties  low-ionization 
lines directly to the relatively stable near-UV and visible spectrum 
of the central object.  However, collisional excitation at some 
reasonable temperature is still needed to populate the low-energy 
metastable states,  facilitating  continuum pumping to higher states. 
As noted earlier, low metastable states of Fe$^+$, Ti$^+$, etc.,
in the blobs and Sr Filament appear to have LTE-like populations 
with $T \sim 6000$ to 7000 K  (\S4.2, and \cite{bau02}).  
Somehow this occurs without much hydrogen ionization.  
This presents a problem because at 6000--7000 K with 10$^7$ electrons per
   cm$^{3}$,  the Saha eqn predicts  more  H$^+$ than  H$^0$!

\section{Summary: The Nature and Origin of the Inner Ejecta}  

The inner ejecta are dominated observationally by the Weigelt blobs, 
which appear to be concentrations of warm, relatively dense 
gas that is heated, photo-excited and usually (apart from the 
spectroscopic events) photo-ionized by the central continuum source. 
Considerable amounts of gas also exist outside the blobs, including 
prominent absorption line regions and some faster and less 
dense gas that emits [Ne~III], [Ar~III], etc.  The blobs, at least, 
represent heavily CNO processed gas emitted from the primary   
star roughly a century ago. There appears to be little  
dust within the main emitting and absorbing condensations of the 
inner ejecta, but our knowledge  of the amount and spatial distribution 
of this material is limited by uncertainties in the patchiness and 
amount of the (rather gray) foreground extinction.

There are at least two leading unsolved puzzles in the emission 
line physics. The most fundamental concerns the heating 
and weak ionization of the gas that produces strong 
emission from ions like Fe$^+$, Ni$^+$, Ca$^+$, Ti$^+$ and 
Sr$^+$.  Existing photoionization and photo-excitation models 
imply that this gas should be partially ionized, with significant 
amounts of H$^+$;  but the data indicate that hydrogen is practically 
non-ionized in the blobs during the spectroscopic events (and in the Sr 
filament at all times).  Another puzzle 
involves the bizarre line ratios emitted from Ly$\alpha$-pumped 
levels of Fe$^+$ that produce the strong \lam 2508 and
 \lam 2509 lines. 
  
  We conclude with a reminder about the broader goal of studies of the 
  inner ejecta -- to understand the nature and evolution of the central 
  object.  Here let us mention one particular topic that deserves   
  more study. During the transitional phases at the beginning and end 
  of a spectroscopic event, the UV flux from the central star(s) is 
  extinguished by varying amounts/properties of a shielding gas. Moreover, 
  the event timings observed in various ions (\S3) imply that the 
  cutoff energy slides from the far-UV to the near UV and back again. 
  Calculations are needed to see what might cause this behavior. 
  It cannot occur merely by variable column densities in a neutral 
  medium. A more realistic scenario would involve column densities 
  in partially ionized gas, possibly combined with varying degrees 
  of ionization. 

In the binary model of the central object, radiative 
shielding occurs when the hot companion star plunges deep 
inside the dense wind of the primary (\S2). The shielding medium 
is the dense wind, perturbed or enhanced by its interaction with 
the hot binary. Quantitative spectral studies of the inner 
ejecta, especially during the transition phases, should be very 
helpful for constraining basic properties of the companion star, 
its wind, and the binary orbit/orientation.  For a few days 
before and after periastron passage, some regions of the inner 
ejecta should be lit up by far more UV radiation than others. 
New observations with enough spatial and temporal resolution 
might allow us to see this pattern of illumination move across the 
inner ejecta, as has been suggested already for some nebulosity 
farther out in the Homunculus \cite{s+04ph}. The best tracer for these 
effects is probably the [Ne~III] \lam 3868 line because its 
emission is tied directly to the far-UV output from the hot 
companion and its profile is not blended with reflected 
features from the stellar wind (see \cite{meh10a}).

\begin{acknowledgement} FH  grateful to the HST--Eta Carinae 
Treasury Team, especially Kris Davidson and Bish Ishibashi, for their 
help and guidance with the HST spectra. Brian Cherinka also helped 
with some of the data processing. FH  had valuable discussions about 
nebular physics with Gary Ferland and Pat Hall.  Andrea Mehner 
contributed recent information, especially for {\S}4.1.  Finally, 
I thank the editors Roberta Humphreys and Kris Davidson for useful 
comments. 

\end{acknowledgement}

\end{document}